\documentclass[preprint]{aastex}

\usepackage{natbib}
\usepackage{graphicx}
\usepackage{xcolor}
\usepackage{rotating}
\usepackage{epsfig}
\usepackage{epstopdf}
\usepackage{enumitem}
\usepackage{amssymb,amsmath}
\usepackage{longtable}
\usepackage{subfig}

\newcommand{\Rs}{$ R_{\odot}$}

\newcommand{\de}{$^{\circ}$}
\newcommand{\deday}{$^{\circ}$/day}%$^{-1}$}

\newcommand{\kms}{~km\,$\text{s}^{-1}$}

\newcommand{\cmvol}{cm$^{-3}$}
\newcommand{\paperi}{Paper I}
\newcommand{\paperii}{Paper II}
\newcommand{\paperiii}{Paper III}
\newcommand{\app}{$\approx$}
\makeatletter

\newcommand{\Rmnum}[1]{\expandafter\@slowromancap\romannumeral #1@}

\makeatother

\begin{document}

\shorttitle{Solar coronal rotation}
\shortauthors{Morgan \& Edwards}
\title{A solar-cycle study of coronal rotation:
large variations, rapid changes, and implications for solar wind models}
\author{Liam Edwards} 
\author{David Kuridze} 
\author{Thomas Williams} 
\author{Huw Morgan} 
\affil{Department of Physics, Aberystwyth University, Ceredigion, Cymru, SY23 3BZ, UK}
\email{hmorgan@aber.ac.uk}

%%%%%%%%%%%%%%%%%%%%%%%%%%%%%%%%%%%%%%%%%%%%%%%%%%%%%%%%%%%%%%%%%%%%%%%%%%%%%%%%%%%%%%%%%%%%%%%%%%%%%%%%%%%%%%%%%%%%%%%%%%
\begin{abstract}
Information on the rotation rate of the corona, and its variation over latitude and solar cycle, is valuable for making global connections between the corona and the Sun, for global estimates of reconnection rates, and as a basic parameter for solar wind modelling. Here, we use a time series of tomographical maps gained from coronagraph observations between 2007\,--\,2020 to directly measure the longitudinal drift of high-density streamers over time. The method reveals abrupt changes in rotation rates, revealing a complex relationship between the coronal rotation and the underlying photosphere. The majority of rates are between -1.0 to +0.5\deday\ relative to the standard Carrington rate of 14.18\deday, although rates are measured as low as -2.2\deday\ and as high as 1.6\deday. Equatorial rotation rates during the 2008 solar minimum are slightly faster than the Carrington rate, with an abrupt switch to slow rotation in 2009, then a return to faster rates in 2017. Abrupt changes and large variations in rates are seen at all latitudes. Comparison with a magnetic model suggests that periods of equatorial fast rotation are associated with times when a large proportion of the magnetic footpoints of equatorial streamers are near the equator, and we interpret the abrupt changes in terms of the latitudinal distribution of the streamer photospheric footpoints. The coronal rotation rate is a key parameter for solar wind models, and variations of up to a degree per day or more can lead to large systematic errors over forecasting periods of longer than a few days. The approach described in this paper gives corrected values that can form a part of future forecasting efforts.
\end{abstract}

\maketitle

%%%%%%%%%%%%%% INTRODUCTION %%%%%%%%%%%%%%%%%%%%%%%%%%%%%

\section{Introduction}
\label{intro}
The rotation of the Sun is a complicated subject due to the variation of results arising from a broad range of observations and analysis techniques. The variation of results is a symptom of the complexity of the Sun's rotation arising from the interdependence of the solar magnetic dynamo with the convective plasma, leading to a rotation which is dependent on latitude, depth within the Sun, and solar cycle phase. See \citet{howard1984}, \citet{schroeter1985}, \citet{thompson2003}, and \citet{beck2000} for reviews. The rotation of the solar atmospheric layers, in relation to the underlying photosphere, gives insight into how different atmospheric structures (e.g. active regions, coronal holes) may be influenced by sub-photospheric motions, and to rates of interchange reconnection in the low atmosphere. It is a field made complicated by many factors including the different dependence of types of observation on density and geometrical factors (e.g. radio, white light, line emission), the temperature dependence of different observations, uncertainties on a local level as to the magnetic connectivity between different atmospheric layers, lack of direct routine observations of the atmospheric magnetic field, and above all, the extended line of sight through the optically thin medium. 

The rotation of the optically thin corona can be estimated using several different types of observations. For heights above the very lowest corona (or where the atmospheric features cannot be observed against the disk, necessitating the use of off-limb observations), most rotation estimates are based on long time-series, subject to an extended line of sight (LOS). As different coronal structures rotate through the field of view, the signal's modulation gives an estimate of the dominant rotation rate. This is commonly called flux modulation, and it must use long time series to determine the rotation rate (several months to years), thus shorter time-scales are inaccessible. We refer the reader to the introduction of \citet{morgan2011rotation} for a more detailed overview of solar atmospheric rotation. 

There have been several works studying coronal rotation since \citet{morgan2011rotation}. \citet{vats2011} used long-term flux modulation of both radio and X-ray imagery to show a small yet clear differential rotation of the corona, with a north-south asymmetry in rotation rates, with asymmetry most pronounced during solar maximum. A similar asymmetry has been shown using extreme ultraviolet (EUV) observations by \citet{sharma2020}. \citet{li2012} used a very long time series of solar radio observations to show a small decreasing trend in the mean coronal rotation rate between years 1947 to 2009, but no significant link to the Schwabe cycle. \citet{xie2017} also found a decreasing trend over the same period, and found significant periodicities in the temporal variation of rotation rates ranging from 2 to 10 years. Conversely, the study of the coronal green line by \citet{deng2020} showed an increasing rotation rate over a similar multi-decadal period, finding similar significant periodicities ranging from 3 to 11 years, and thus a possible link between rotation rates, the 11-year solar cycle, and the quasi-biennial oscillation. Using radio observations, \citet{bhatt2017} found a decreasing rotation rate with greater altitude, in disagreement with both radio measurements by \citet{vats2001}, and EUV measurements by \citet{sharma2020b}, who found an increasing rate with altitude. \citet{obridko2020} used potential field source surface magnetic models of the corona, based on photospheric magnetic field estimates, to study differential rotation as a function of latitude and height, and interpreted their results in terms of links to subphotospheric rotations. Their interpretation is interesting, but does not tie in the large body of results gained from decades of observation. Mechanisms must be invoked and proven (e.g. interchange reconnection) to properly link the coronal rotation and large-scale coronal structural changes to photospheric and subphotospheric rotation and bulk motions. The large body of observational results relating to coronal rotation and structure, as outlined here, in \citet{morgan2011rotation}, and elsewhere, contain results that are often contradictory, and consolidation is required for further understanding. One missing aspect is the use of long-term simulations to gauge the response of the coronal rotation to the changing distribution of the photospheric field - global magnetofrictional models may be ideally suited for this.

\citet{morgan2011rotation} first used tomography maps of the corona to estimate coronal rotation rates at a height of 4\,\Rs\ over the period 1996 to 2010, and found large variations of rotation rates compared to other studies, up to $\pm3$\deday\ relative to the Carrington rate. The study found surprising variations in rotation between latitudes, and rapid changes in rotation rate at a given latitude. A related work also used a time series of tomography maps to study the longitudinal drift of density structures in the equatorial streamer belt during the 1996 solar minimum \citep{morgan2011longitudinaldrifts}. The approach of \citet{morgan2011rotation} differs considerably from a flux modulation approach, since a reconstruction of the coronal density structure is made using two weeks of input data, and the rotation is estimated from the changing configuration of the reconstructed densities. So whilst flux modulation gives the dominant mean rotation rate from a time series spanning at least a few full rotations (several months), a time series of tomography maps give a more direct measurement of rotation with finer time resolution. Projection effects, and the related limitations of flux modulation, are examined by \citet{mancuso2013} in the context of ultraviolet emission line measurements. In particular, they show that projection effects can lead to a bias towards finding a rigid rotation at higher latitudes. The same arguments apply equally to any study of rotation based on flux modulation, thus the combination of projection effects, and long time-series averaging, may lead to a misleading picture of rigid coronal rotation.

This current study extends on the work of \citet{morgan2011rotation} by studying the coronal rotation for years 2007 to 2020. Here we use a much-improved tomography method to gain the coronal density structure, and the drift of streamers in longitude are traced manually, thus avoiding uncertainties or ambiguities associated with automated methods. The observations, tomography method, and the manual method for measuring rotation are summarised in section \ref{method} (an automated correlation-based method is described in appendix \ref{app1}) with the results presented in section \ref{results}. We also provide interpretation, based on magnetic models, of what may be causing abrupt changes in rotation rates in section \ref{magmodel}, and briefly describe the relevance to solar wind models in section \ref{solarwind}. Conclusions are given in section \ref{conclusions}.

%%%%%%%%%%%%%% METHOD %%%%%%%%%%%%%%%%%%%%%%%%%%%%%

\section{Method}
\label{method}

\citet[][hereafter \paperii,]{morgan2019} presents a recent advancement in coronal rotational tomography which gives maps of the coronal electron density at heliocentric distances greater than \app\,3\,\Rs, at all periods of the solar cycle. The method is based on a spherical harmonic model of the coronal density, constrained by coronagraph data that are pre-processed and calibrated as described by \citet[][hereafter \paperi]{morgan2015}. Initial results and further method developments are presented in \citet[][hereafter \paperiii]{morgan2020}. The resulting maps clearly show the large-scale distribution of the streamer belt, albeit showing structure that is smoother than the true density distribution. The maps are based on a static reconstruction, or the smooth and positive distribution of density that best satisfies two weeks of input data. 

The COR2 coronagraphs are part of the Sun Earth Connection Coronal and Heliospheric Investigation (SECCHI, \citet{howard2002}) suite of instruments aboard the twin Solar Terrestial Relations Observatory (STEREO A \& B, \citet{kaiser2005}). Half a solar rotation (or \app\,2 weeks) of COR2A observations are needed to create an electron density map using the calibration processes of \paperi, the tomographic inversion of \paperii, and the refinement methods of \paperiii. Using the facilities of SuperComputing Wales, for data spanning 2007 March 17 to 2019 September 5, tomography maps have been created at \app\,2 day increments, resulting in over 2000 sets of maps over the period. The period includes the 2014 to 2015 data gap when STEREO A was traversing behind the Sun. Since each map at a given date is a static reconstruction created from $\pm$1 week of data from that date, the time series of maps over several years can be considered as a `sliding window' of reconstructions. For each date, a set of 9 maps are created for heliocentric distances between 4\,--\,8\,\Rs\ at 0.5\,\Rs\ increments: this work uses the 4\,\Rs\ maps only.

\begin{figure}
    \centering
    \includegraphics[width=0.55\textwidth]{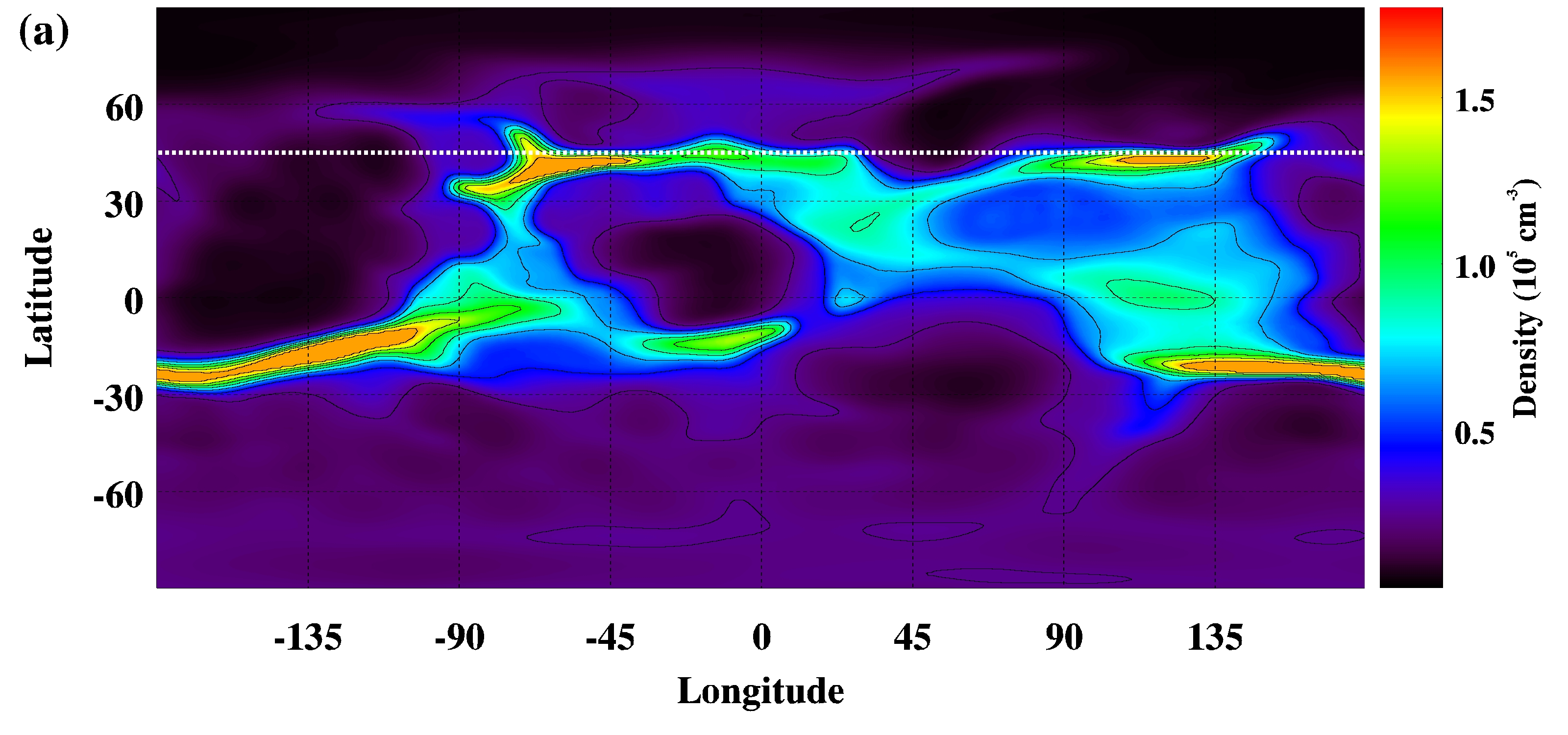}
    \includegraphics[width=0.55\textwidth]{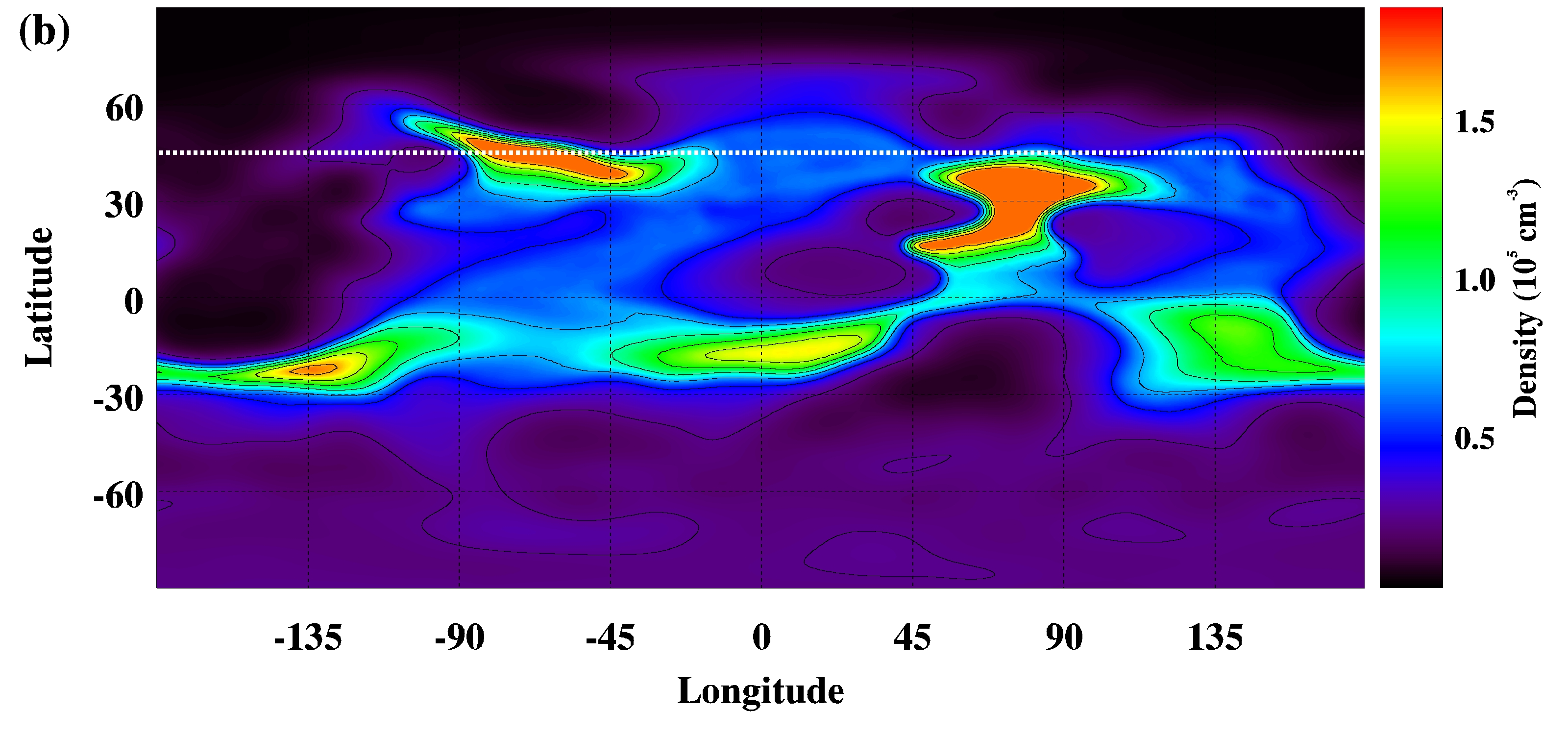}
    \includegraphics[width=0.55\textwidth]{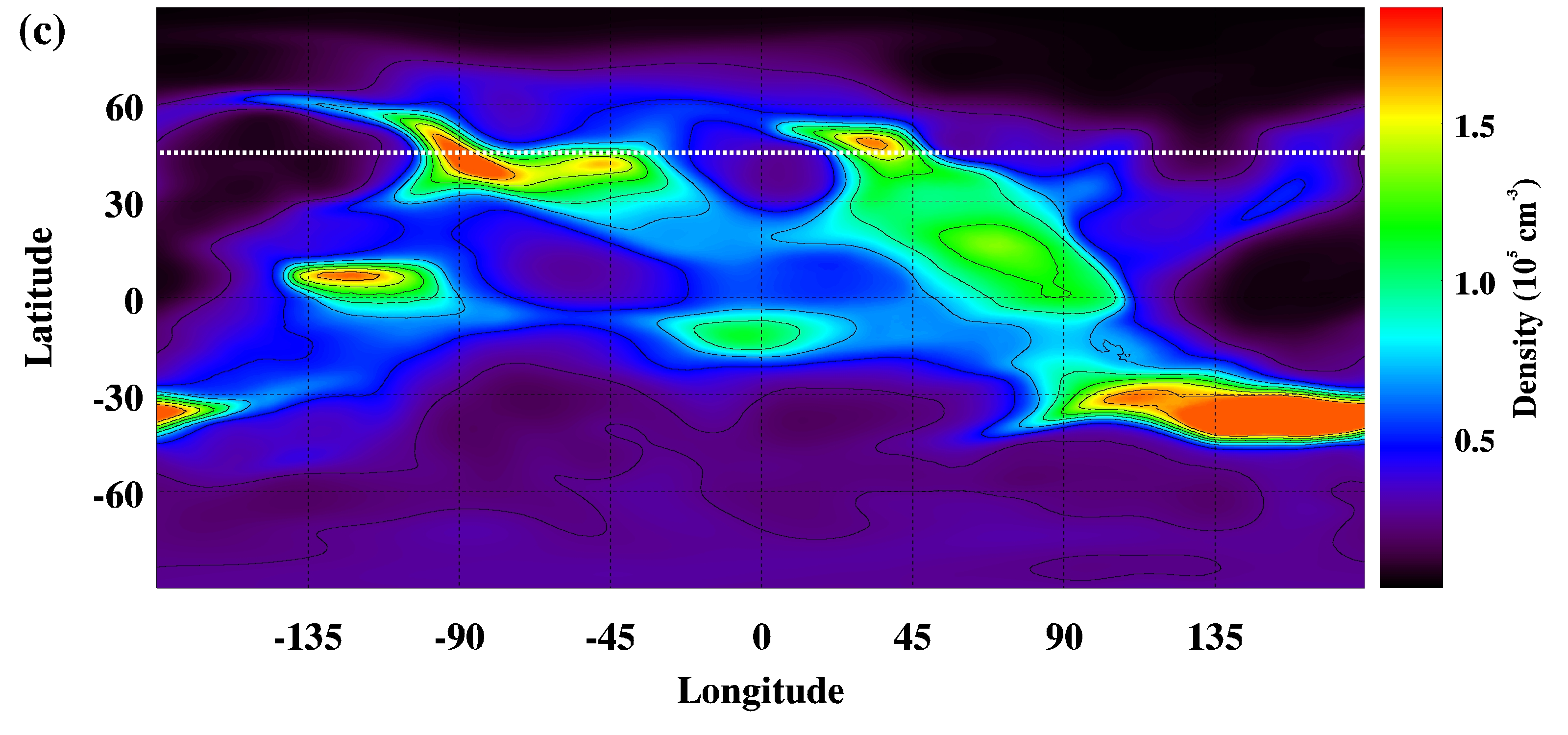}
    \includegraphics[width=0.55\textwidth]{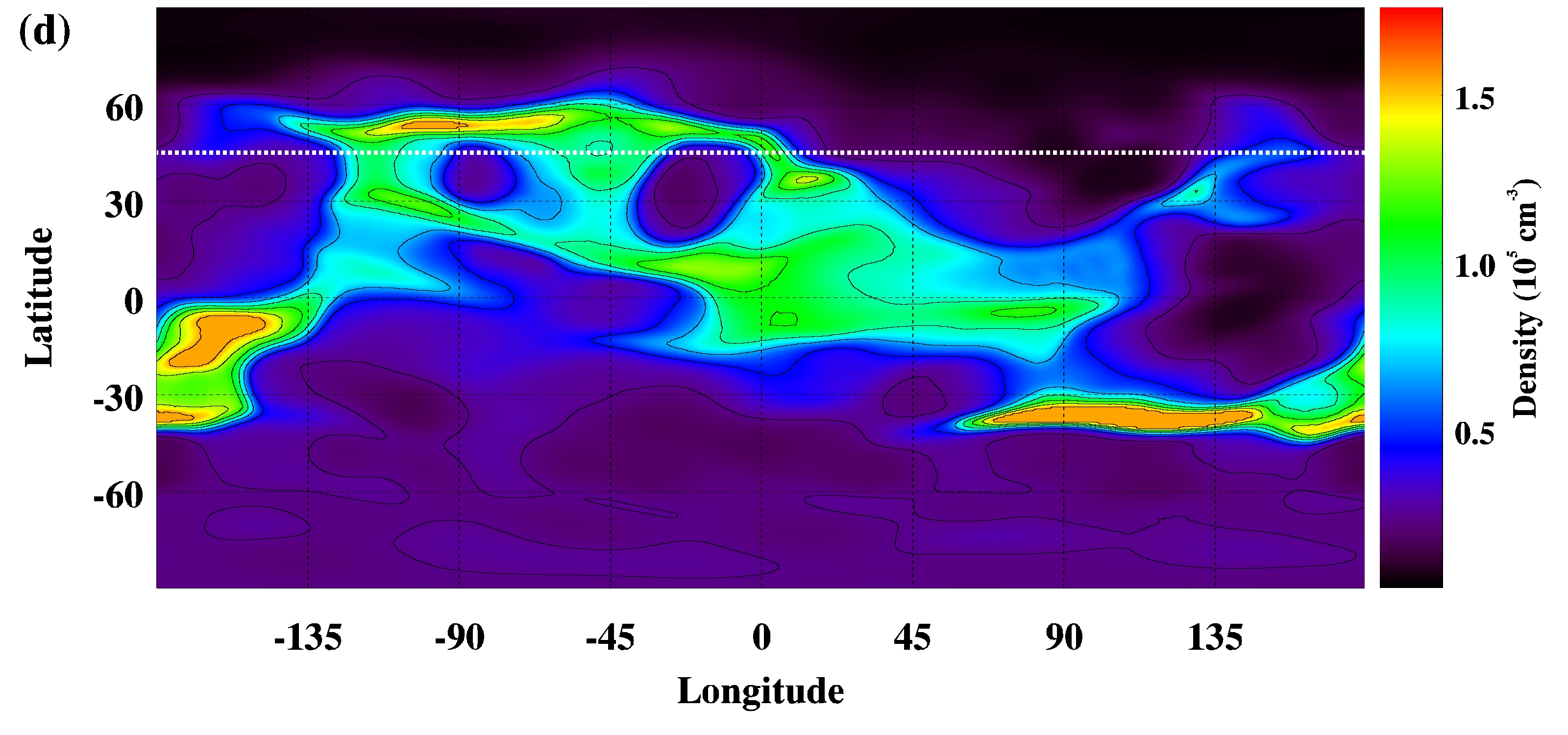}
    \caption{Maps of the coronal electron density at a distance of 4\,\Rs\ for dates (a) 2011 February 3, (b) 2011 March 2, (c) 2011 April 2, and (d) 2011 May 2. The longitude and latitude are Carrington spherical co-ordinates. The density is as given in the colour bars, in units of $10^5$\,\cmvol, with all maps sharing a common colour scale. The horizontal (constant latitude) dashed white line is at a latitude of 40\de, and illustrates how a latitudinal slice can be extracted from the density map in order to analyse the coronal rotation.}
    \label{tommaps}
\end{figure}

Figure \ref{tommaps} shows an example of density maps for four dates in early 2011. When a time series of such maps are viewed, it is common to see a drift of the high-density streamers in longitude, as well as other changes in structure. For example, for Figure \ref{tommaps} from the earliest map at 2011 February 3, through to 2011 May 2, the high-density streamer structure drifts generally to lower longitudes. This simple example shows how we can extract densities at constant latitudes from the maps in order to analyse the coronal rotation. 

\begin{figure}
    \centering
    \includegraphics[width=0.48\textwidth]{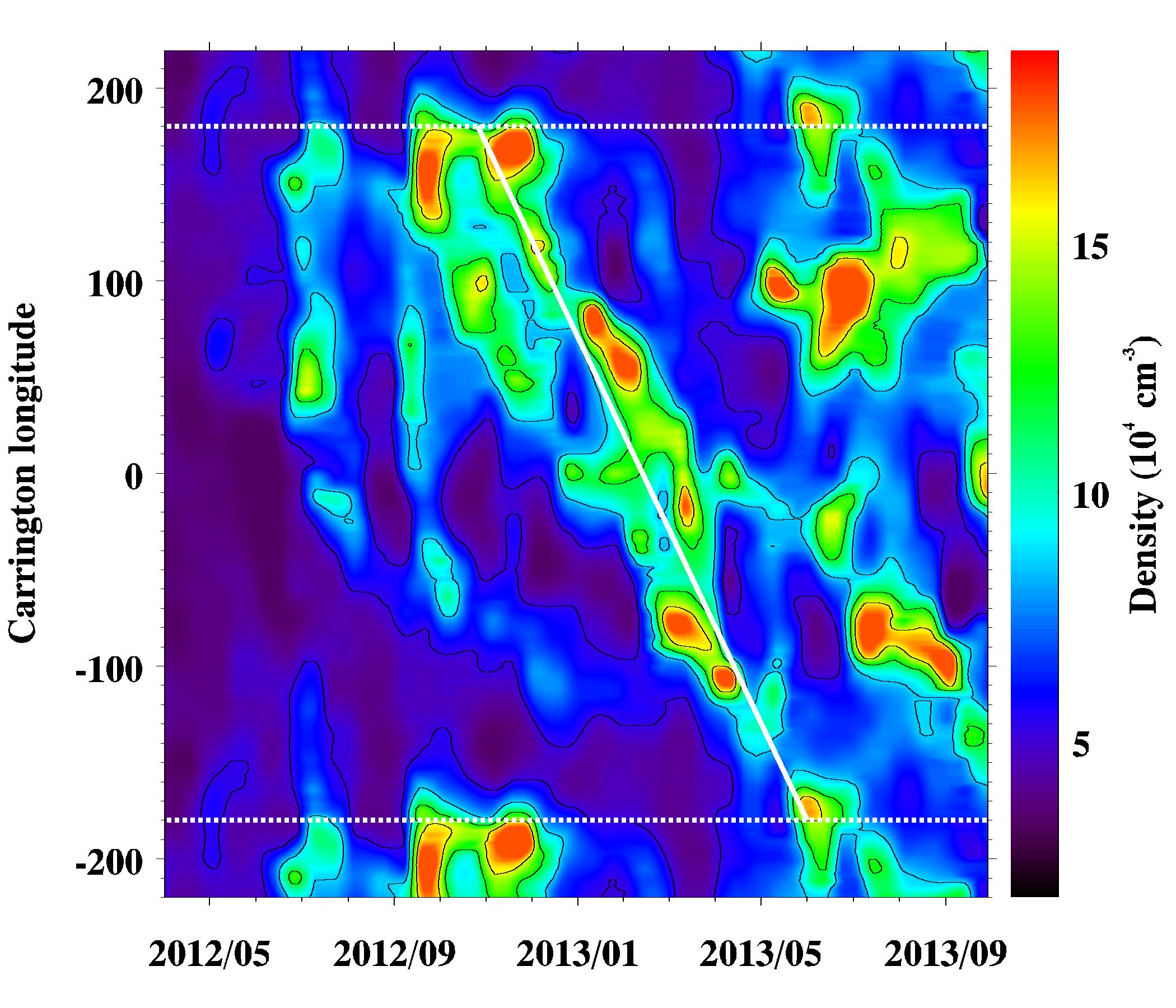}
    \caption{An example of density as a function of time and Carrington longitude, with density indicated by the colour bar. One clear example of a streamer drifting in longitude over time is traced by the solid white line. Note that the longitude axis is wrapped beyond the $\pm180$\de\ in order to show continuation, thus the same information is repeated at the bottom and top of the plot. The $\pm180$\de\ lines are shown by the white dashed lines.}
    \label{grad}
\end{figure}

Profiles of density at a given latitude are stacked in time, giving a time-longitude array of densities. Figure \ref{grad} shows an example time-longitude map for the \app\,2 year period beginning on 2012 April, at a latitude of -60\de. Density structures that rotate at the Carrington rate appear as horizontal features in these plots. This example shows, from the end of 2012 to the end of 2013, one high-density streamer that has a consistent and linear drift with a negative gradient in Carrington longitude, meaning the streamer is rotating slower than the Carrington rate. Across the whole 12 years of data, and across all latitudes, there are numerous examples of positive and negative drifts. For each clear example of a coherent drifting streamer, we manually trace a straight line, as shown in Figure \ref{grad}. The gradient is calculated, which gives the mean rotation rate of the streamer relative to Carrington rotation. For the example of Figure \ref{grad}, we have a gradient of $\frac{\Delta y}{\Delta x} = \frac{-360}{220} = -1.64 \text{\deday}$. Manually identifying and tracing these broad and varying features involves uncertainty. For example, the streamers only drift slowly and sometimes contain multiple density peaks at the same longitude over time (as can be seen at the start of the solid white line in Figure \ref{grad}). Furthermore, the path described by the streamer over time may not follow an exact straight line, which is also seen in the example. Quantifying the uncertainty is difficult. Given approximate time uncertainty of $\pm$30 days and a longitudinal uncertainty of $\pm$10\de, for a streamer that drifts through 360\de\ over one year, the uncertainty in the resulting rotation rate is around 15\%. 

\begin{figure*}
    \centering
    \includegraphics[width=0.8\textwidth]{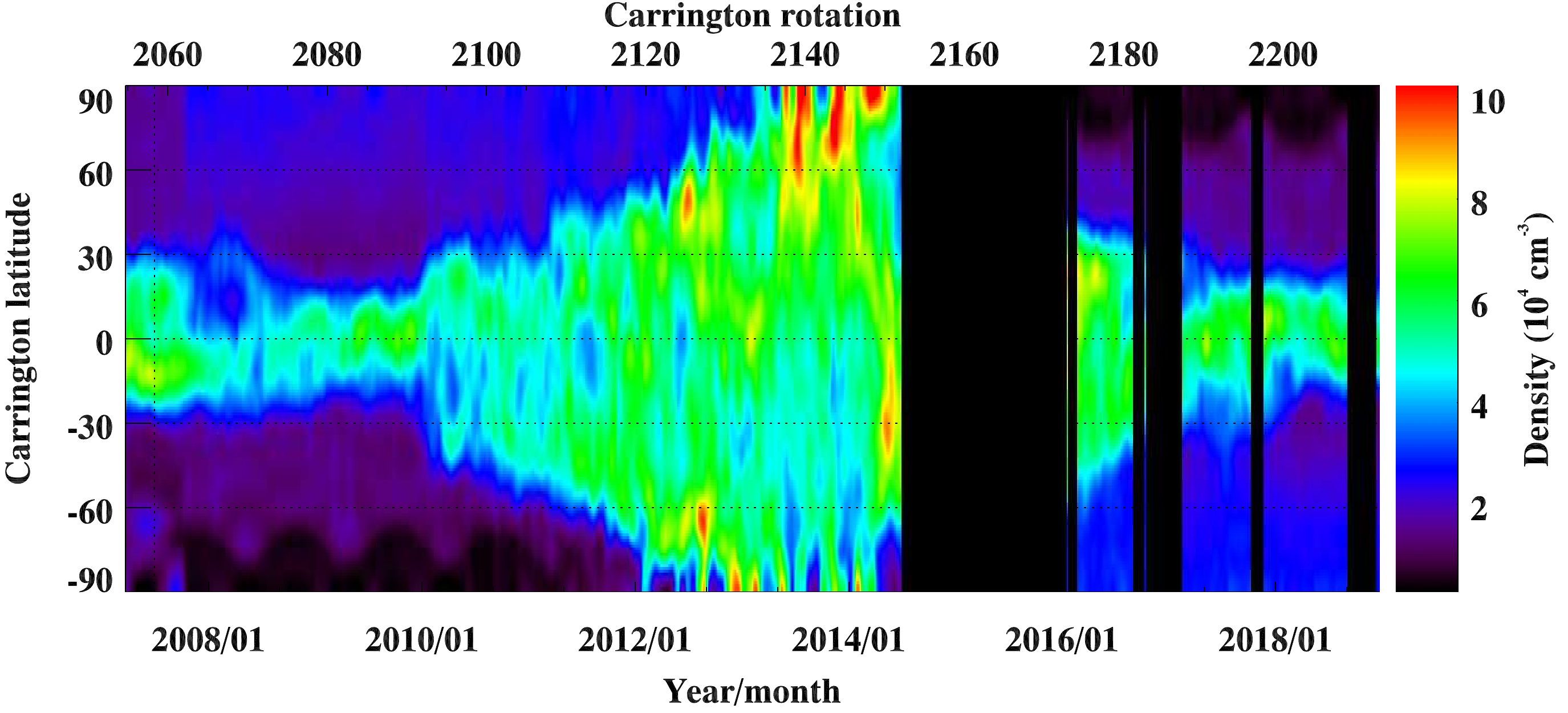}
    \caption{Density, averaged over all longitudes, as a function of latitude and time. The vertical black blocks are either datagaps, periods where data is too scarce to apply tomography, or periods where the tomography has failed to reconstruct.}
    \label{latdist}
\end{figure*}

The solar cycle latitudinal distribution of streamers is shown in Figure \ref{latdist}, which shows the density averaged over all longitudes as a function of time and latitude. This distribution shows that a full 12-year analysis at all latitudes is not possible. So at increasing latitudes from the equator, the study becomes increasingly limited to the years surrounding solar maximum, when high-density streamers exist at higher latitudes. 

%%%%%%%%%%%%%% RESULTS %%%%%%%%%%%%%%%%%%%%%%%%%%%%%

\section{Results}
\label{results}

\begin{figure*}
    \centering
    \includegraphics[width=0.95\textwidth]{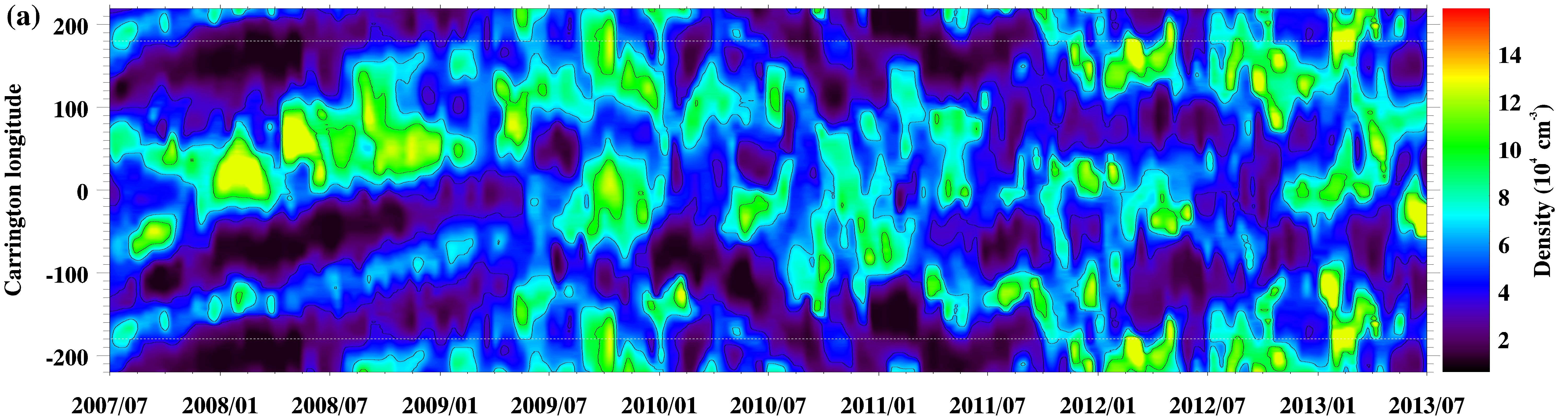}
    \includegraphics[width=0.95\textwidth]{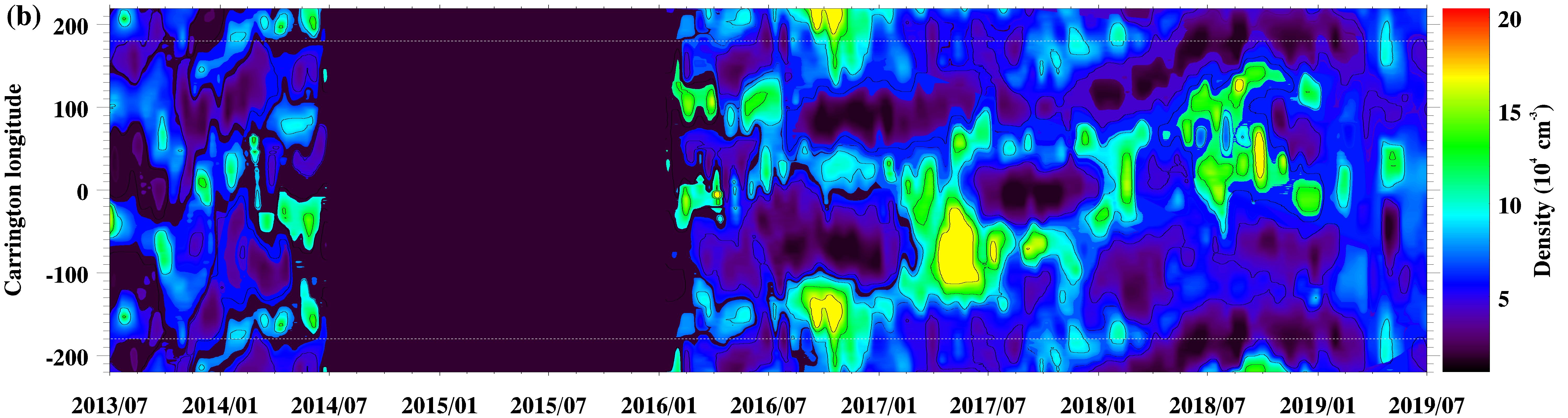}
    \caption{Density in the equatorial plane at a height of 4\,\Rs, as a function of time and Carrington longitude for dates (a) 2007 July 1 to 2013 July 1, and (b) 2013 July 1 to 2019 July 1. There is a long datagap from 2014 July to 2016 January. Interpolation is used to fill shorter datagaps (the times of these shorter datagaps can be seen in Figure \ref{latdist}).}
    \label{equatorial}
\end{figure*}

Figure \ref{equatorial} shows an example time-longitude density plot for (a) the solar minimum and ascending phase six-year period starting on 2007 July 1, and (b) the following six-year period including solar maximum and the descending phase. This latter period also contains a long period of missing tomography data from $\approx$\,2014 July to 2016 January caused by telemetry disruption as STEREO A passed behind the Sun. The figure shows a clear rotation rate approximately 0.25\deday\ faster than the Carrington rate from 2007 January to 2009 October, followed by a two year period of rotation considerably slower than the Carrington rate. During the years surrounding solar maximum, there are less clear signatures of rotation rates, and the large-scale structures are short-lived compared to solar minimum and the ascending phase; lasting months rather than years. Despite this, some of these can be traced over time to calculate rates. From mid-2016 onwards the rotation returns to a rate faster than the Carrington rate, similar to the 2008 solar minimum rate. We attempt to explain the abrupt change in rate during 2009 October by the connection between coronal streamers and the underlying lowest corona and photosphere in the discussion.

\begin{figure}
    \centering
    \includegraphics[width=0.49\textwidth]{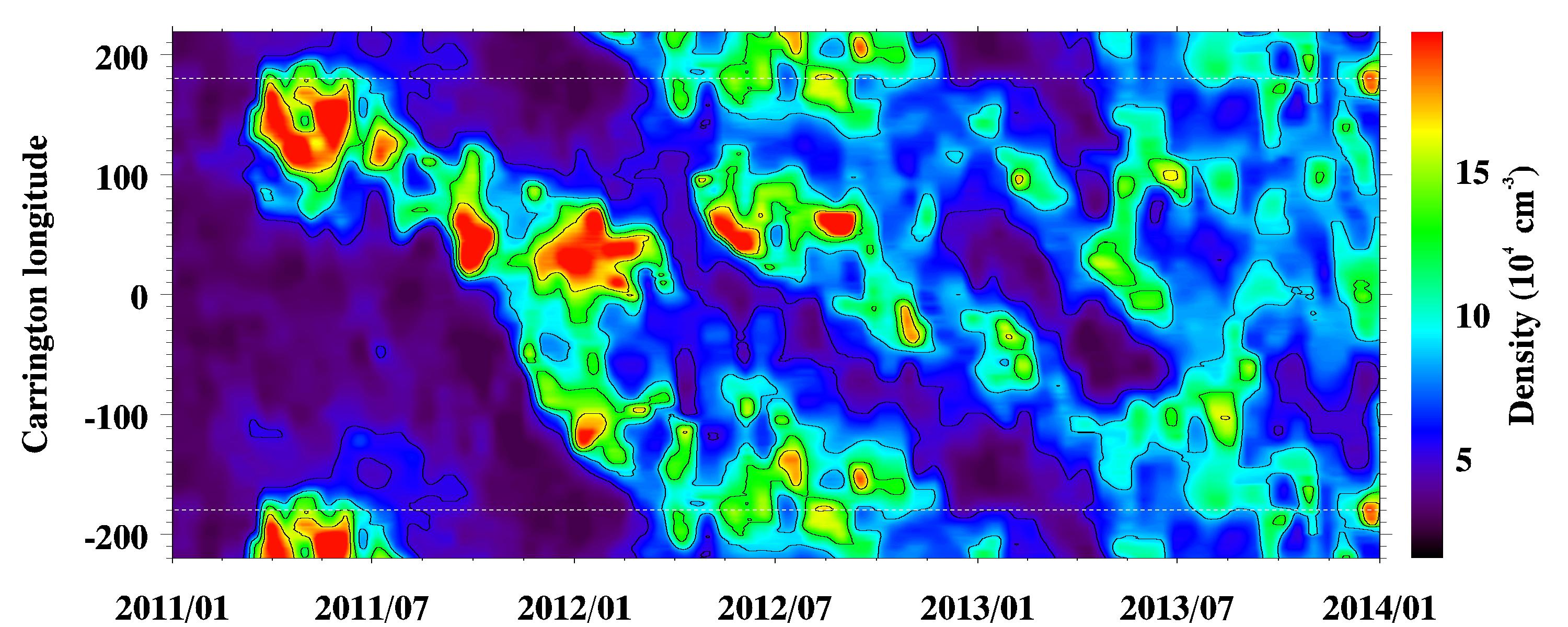}
    \caption{Density at the south mid-latitude ($-50$\de) for 2011 January to 2014 January.}
    \label{midlatdens}
\end{figure}

\begin{figure}
    \centering
    \includegraphics[width=0.49\textwidth]{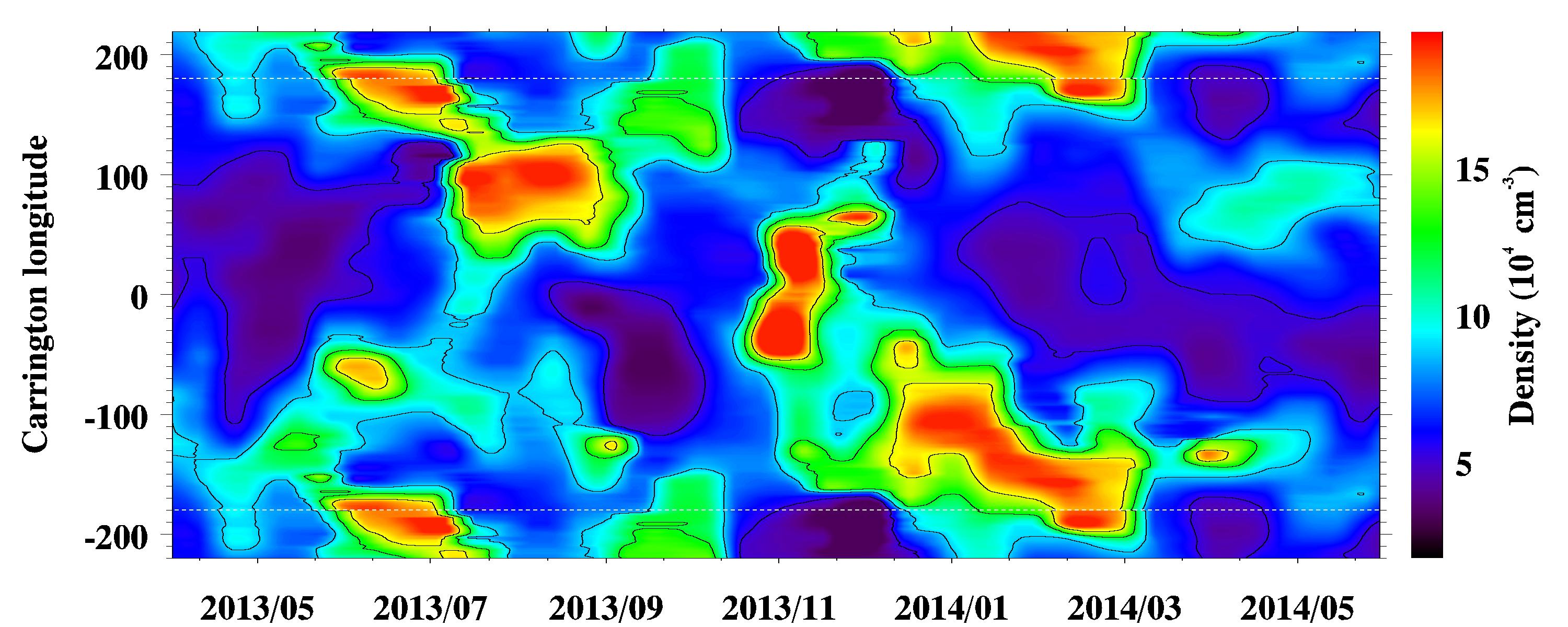}
    \caption{Density at high south latitude ($-80$\de) for 2013 April to 2014 June.}
    \label{polardens}
\end{figure}

Figure \ref{midlatdens} shows the density evolution for latitude $-50$\de\ for the period from when streamers first appear at this latitude (early 2011) through to 2014 January. Slow rotation can be seen up to around mid-2013. The streamer distribution then becomes incoherent and uncertain to interpret in terms of rotation. Figure \ref{polardens} shows the density near to the south pole at $-80$\de\ for 2013 April to 2014 June. Although less clear than at lower latitudes, there is a clear negative gradient drift of the main streamer.

\begin{figure*}
    \centering
    \includegraphics[width=0.9\textwidth]{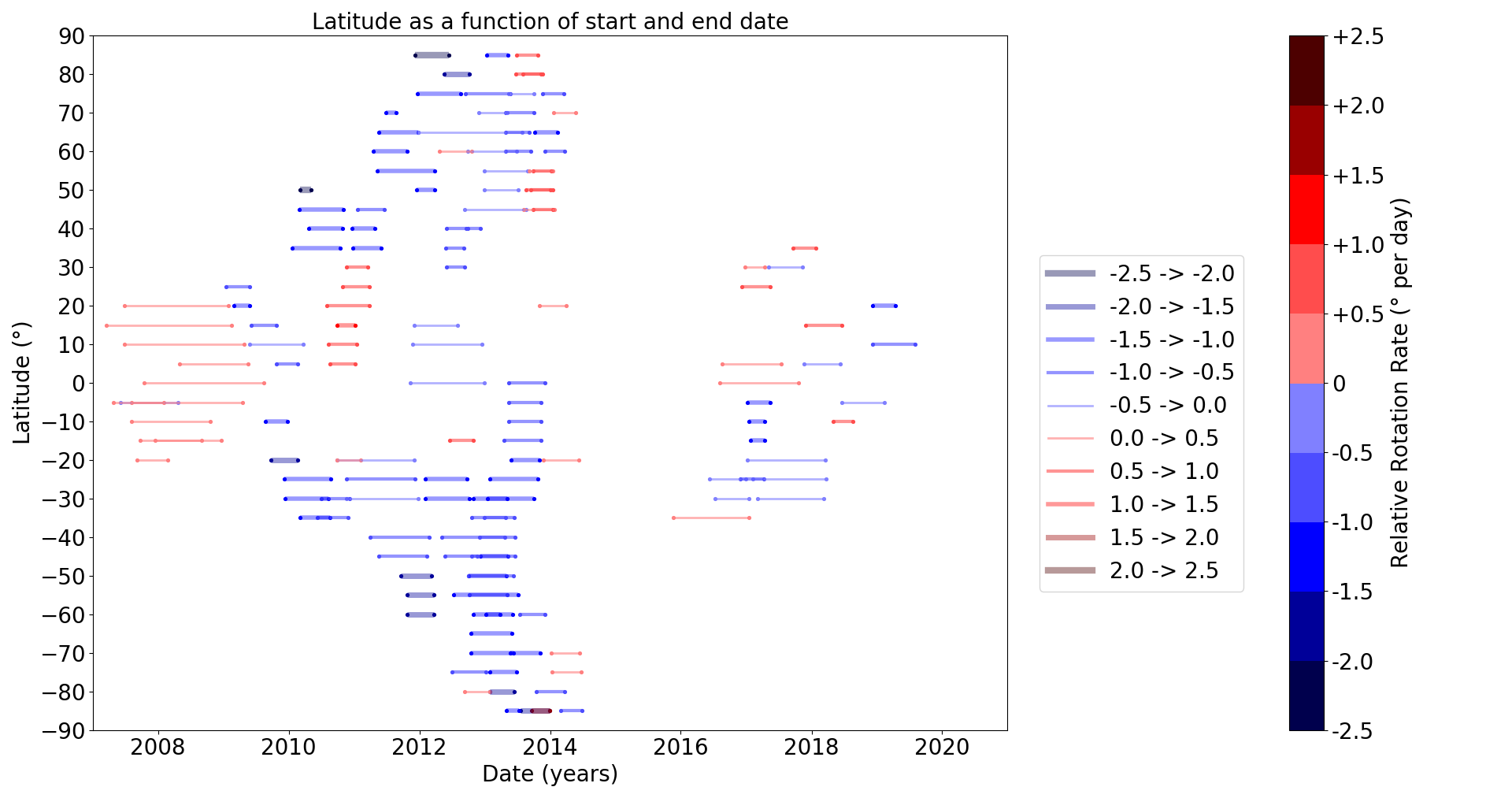}
    \includegraphics[width=0.9\textwidth]{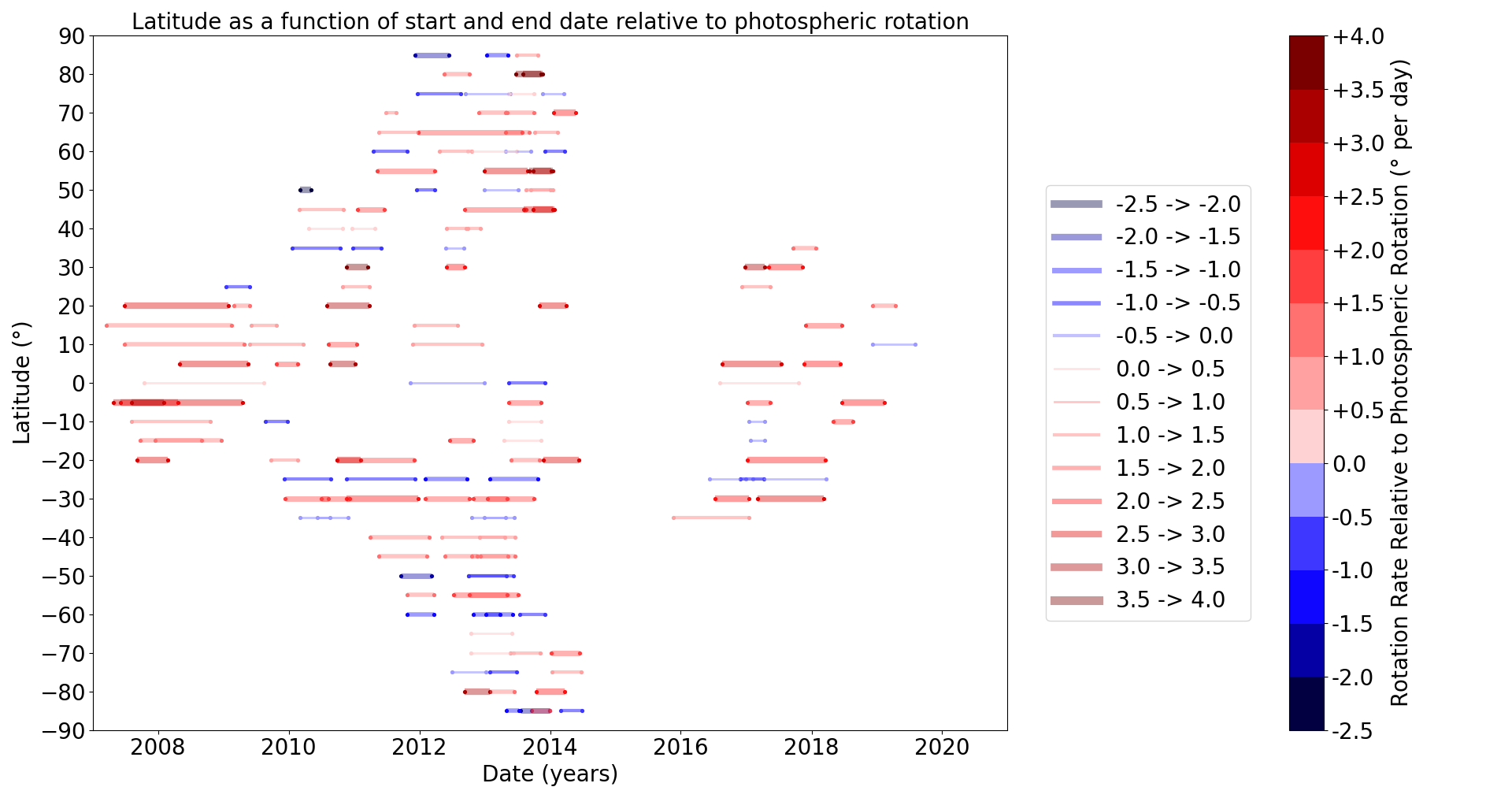}
    \caption{\emph{Top}: rotation rates for coronal streamers as a function of time and latitude. \emph{Red} (\emph{blue}) colours represent faster (slower) rotation rates relative to Carrington rotation, as indicated in the color bar. The horizontal span of each line represents the period over which the rotation rate was estimated. \emph{Bottom}: rotation rates for coronal streamers relative to the underlying photospheric rotation rate at the same latitude, with photospheric rotation rates as given by \citet{howard1984}.}
    \label{rotrat}
\end{figure*}

Table \ref{tablesouth} lists information on all manually traced streamers in the south corona throughout the dataset, including the latitude of each structure, the mid-date and duration in days, and the estimated rotation rate relative to the Carrington rate. Table \ref{tablenorth} shows information for northerly and equatorial streamers. The top plot of Figure \ref{rotrat} visualises this information, showing all measured rotation rates over the solar cycle as a function of time and latitude. The majority of values are slower than the Carrington rotation rate. Faster than Carrington rotation is seen near the equator during solar minimum in 2008, from equatorial to mid-latitude regions in the north during the ascending phase to maximum in 2010 to 2012, and in the polar north during solar maximum. There are also some examples of faster than Carrington rates in the polar south during solar maximum. After the recovery of STEREO during the descending phase to minimum (2017 and 2018), there is a mixture of slow and fast rotation. In general over the solar cycle, faster rotations occur more frequently in the north.

\begin{table}
\begin{center}
\small
\caption{Measured rotation rates (in degrees per day, relative to the Carrington rate), for streamers in the south. The duration is given in days.}
\begin{tabular}{ c | c | c | c || c | c | c | c }
{Latitude} & {Mid-date} & {Duration} & {Rot. rate} & {Latitude} & {Mid-date} & {Duration} & {Rot. rate}\\
\hline
-85&2013/06/07&70&-1.36&-85&2013/10/09&160&-1.84\\
-85&2013/11/09&100&1.60&-85&2014/04/28&120&-0.75\\
-80&2012/11/18&140&0.29&-80&2013/04/07&140&-1.50\\
-80&2014/01/02&160&-0.72&-75&2012/10/05&190&-0.63\\
-75&2013/04/13&150&-1.37&-75&2014/04/03&160&0.28\\
-70&2013/02/11&240&-1.35&-70&2013/08/15&170&-1.00\\
-70&2014/03/28&160&0.28&-65&2013/02/05&230&-1.46\\
-60&2012/01/07&150&-1.70&-60&2013/01/11&150&-1.43\\
-60&2013/03/22&150&-1.33&-60&2013/09/23&140&-0.86\\
-55&2012/01/06&150&-1.73&-55&2013/01/05&360&-1.13\\
-55&2013/01/20&210&-0.98&-50&2011/12/14&170&-1.97\\
-50&2013/01/17&210&-1.17&-50&2013/02/06&250&-0.88\\
-45&2011/10/01&270&-0.67&-45&2012/08/21&180&-0.83\\
-45&2013/02/22&150&-1.00&-45&2013/02/17&240&-0.98\\
-40&2011/09/14&330&-0.50&-40&2012/10/28&350&-0.60\\
-40&2013/03/12&200&-0.93&-35&2010/05/27&170&-1.00\\
-35&2010/09/04&170&-0.91&-35&2013/01/21&190&-0.76\\
-35&2013/03/22&170&-0.97&-35&2016/06/19&420&0.18\\
-30&2010/04/12&240&-1.04&-30&2010/09/19&160&-0.97\\
-30&2011/06/06&400&-0.46&-30&2012/06/05&250&-1.10\\
-30&2013/01/31&190&-1.00&-30&2013/05/26&260&-1.21\\
-30&2016/10/12&190&-0.42&-30&2017/09/07&370&-0.04\\
-25&2010/04/15&260&-1.02&-25&2011/05/30&380&-0.59\\
-25&2012/05/29&230&-1.33&-25&2013/06/13&270&-1.19\\
-25&2016/09/20&200&-0.35&-25&2017/02/02&130&-0.77\\
-25&2017/08/31&410&-0.18&-20&2007/11/29&170&0.41\\
-20&2009/12/08&150&-1.73&-20&2010/12/03&130&0.35\\
-20&2011/05/02&430&-0.44&-20&2013/08/14&160&-1.16\\
-20&2014/03/02&200&0.28&-20&2017/08/13&440&-0.09\\
-15&2008/03/13&340&0.09&-15&2008/06/16&370&0.16\\
-15&2012/08/24&130&0.77&-15&2013/07/30&210&-0.90\\
-15&2017/03/06&80&-1.25&-10&2008/03/13&440&0.15\\
-10&2009/10/24&120&-1.38&-10&2013/08/14&180&-0.78\\
-10&2017/03/01&90&-1.33&-10&2018/06/24&110&0.95\\
-5&2007/11/14&320&-0.27&-5&2007/11/04&180&0.42\\
-5&2008/04/22&720&0.17&-5&2013/08/14&180&-0.72\\
-5&2017/03/11&130&-1.08&-5&2018/10/17&240&-0.13\\
\end{tabular}
\label{tablesouth}
\end{center}
\end{table}

\begin{table}
\begin{center}
\small
\caption{Measured rotation rates for streamers at the equator and in the north.}
\begin{tabular}{ c | c | c | c || c | c | c | c }
{Latitude} & {Mid-date} & {Duration} & {Rot. rate} & {Latitude} & {Mid-date} & {Duration} & {Rot. rate}\\
\hline
0&2008/09/14&670&0.22&0&2012/06/05&410&-0.05\\
0&2013/08/24&200&-0.50&0&2017/03/16&440&0.24\\
5&2008/11/08&380&0.17&5&2009/12/23&120&-0.83\\
5&2010/10/29&140&0.61&5&2017/01/30&330&0.27\\
5&2018/03/01&200&-0.18&10&2008/05/27&670&0.23\\
10&2009/10/24&300&-0.25&10&2010/10/29&160&0.69\\
10&2012/06/05&390&-0.14&10&2019/04/05&240&-0.92\\
15&2008/03/03&700&0.22&15&2009/08/15&140&-0.61\\
15&2010/11/18&100&1.30&15&2012/04/01&240&-0.33\\
15&2018/03/11&200&0.68&20&2008/04/12&580&0.16\\
20&2009/04/12&90&-1.00&20&2010/11/28&240&0.69\\
20&2014/01/16&150&0.33&20&2019/02/09&130&-1.19\\
25&2009/03/23&130&-0.85&25&2011/01/12&150&0.90\\
25&2017/02/24&160&0.59&30&2011/01/18&120&0.92\\
30&2012/07/21&100&-0.75&30&2017/02/20&110&0.50\\
30&2017/08/09&190&-0.42&35&2010/06/03&270&-1.26\\
35&2011/03/15&160&-1.16&35&2012/07/17&100&-0.75\\
35&2017/11/23&130&0.69&40&2010/07/27&190&-1.29\\
40&2011/02/22&130&-1.19&40&2012/07/26&110&-0.68\\
40&2012/11/03&70&-0.86&45&2010/07/02&245&-1.08\\
45&2011/04/05&150&-0.50&45&2013/02/28&340&-0.44\\
45&2013/10/31&170&0.50&45&2013/11/20&110&0.64\\
50&2010/04/05&60&-2.42&50&2012/02/04&100&-1.05\\
50&2013/04/04&190&-0.37&50&2013/10/26&140&0.57\\
50&2013/11/15&120&0.58&55&2011/10/18&320&-1.11\\
55&2013/04/30&240&-0.27&55&2013/11/11&130&0.50\\
55&2013/11/16&100&0.65&60&2011/07/22&190&-1.03\\
60&2012/07/21&180&0.44&60&2013/02/11&270&-0.11\\
60&2013/07/06&140&-0.64&60&2014/01/27&110&-0.77\\
65&2011/09/06&220&-1.25&65&2012/10/10&580&-0.20\\
65&2013/07/02&130&-0.73&65&2013/12/09&130&-1.04\\
70&2011/07/27&60&-1.00&70&2013/02/16&160&-0.31\\
70&2013/07/16&160&-0.50&70&2014/03/23&120&0.50\\
75&2012/04/19&240&-1.13&75&2013/01/19&250&-0.50\\
75&2013/07/23&140&-0.18&75&2014/01/19&120&-0.71\\
80&2012/07/27&140&-1.50&80&2013/08/31&140&0.82\\
80&2013/09/25&110&0.77&85&2012/03/10&190&-2.03\\
85&2013/03/10&120&-1.17&85&2013/08/27&120&0.83\\
\end{tabular}
\label{tablenorth}
\end{center}

\end{table}

The bottom plot of Figure \ref{rotrat} shows this same information, but relative to the underlying photospheric rotation rate at the same latitude, as given by \citet{howard1984}. This plot shows that, in general, the corona rotates faster then the photosphere at the same latitude, up to 4\deday\ for some periods. The generally accepted view of coronal rotation is of a rigid rotator compared to the photosphere, with less differential rotation. Based on this view we would expect the corona to rotate faster than the photosphere at high latitudes. Our results shows that this is true at the majority of times, but is an oversimplistic model of coronal rotation. At certain latitudes and times the corona can rotate slower than the photosphere at the same latitude, and at some times even low-latitude streamers can rotate considerably faster than the equatorial photosphere. The discussion gives some further insight into this complicated pattern of rotation in terms of how the streamer belt is magnetically connected to the low atmosphere. Nevertheless, the bottom plot of Figure \ref{rotrat} shows that the corona is more likely to be rotating faster than the underlying photosphere at all latitudes, which lends support for the connection between large-scale coronal structure and subphotospheric motions.

\begin{figure*}
    \centering
    \includegraphics[width=0.6\textwidth]{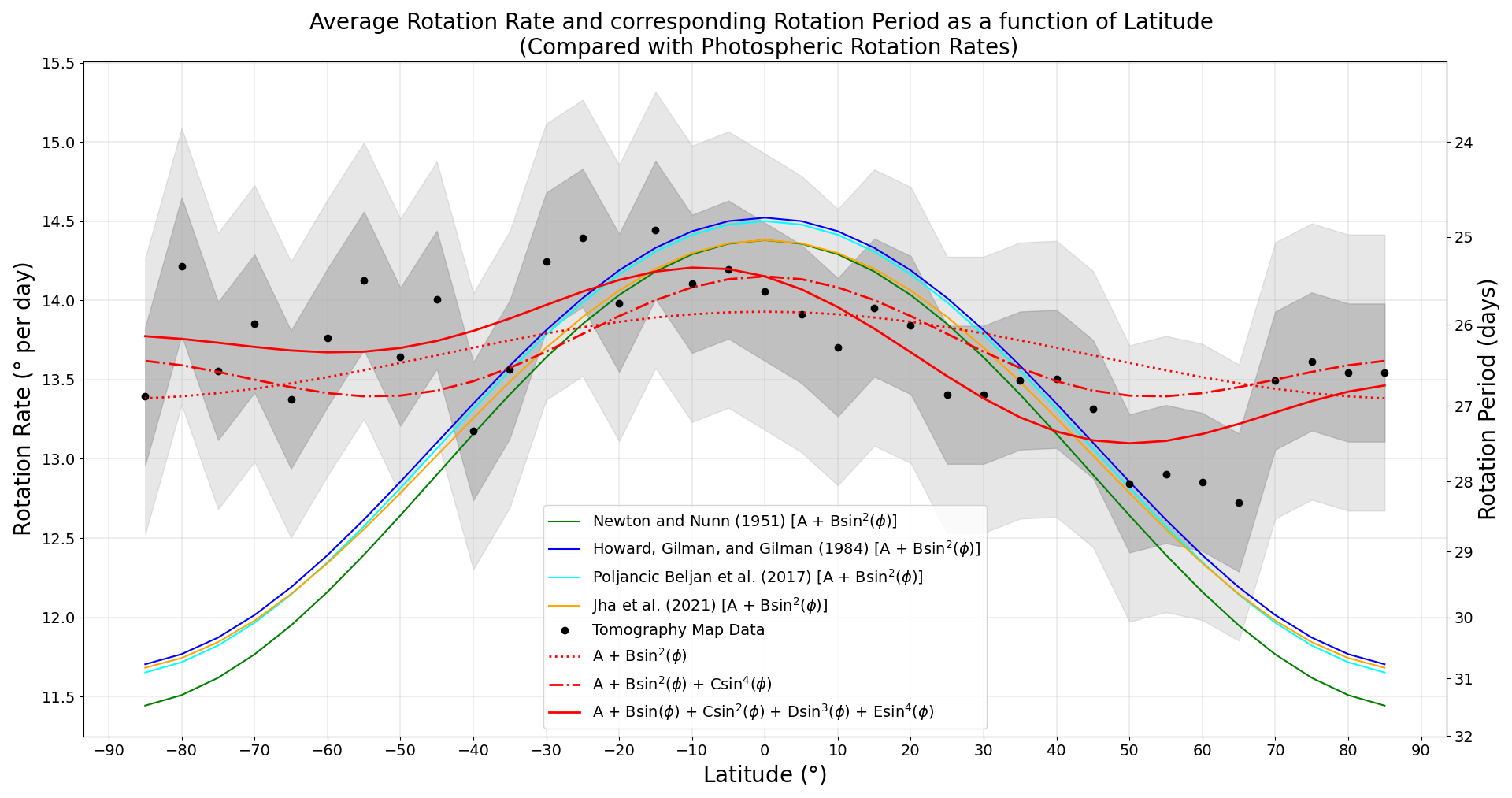}
    \includegraphics[width=0.6\textwidth]{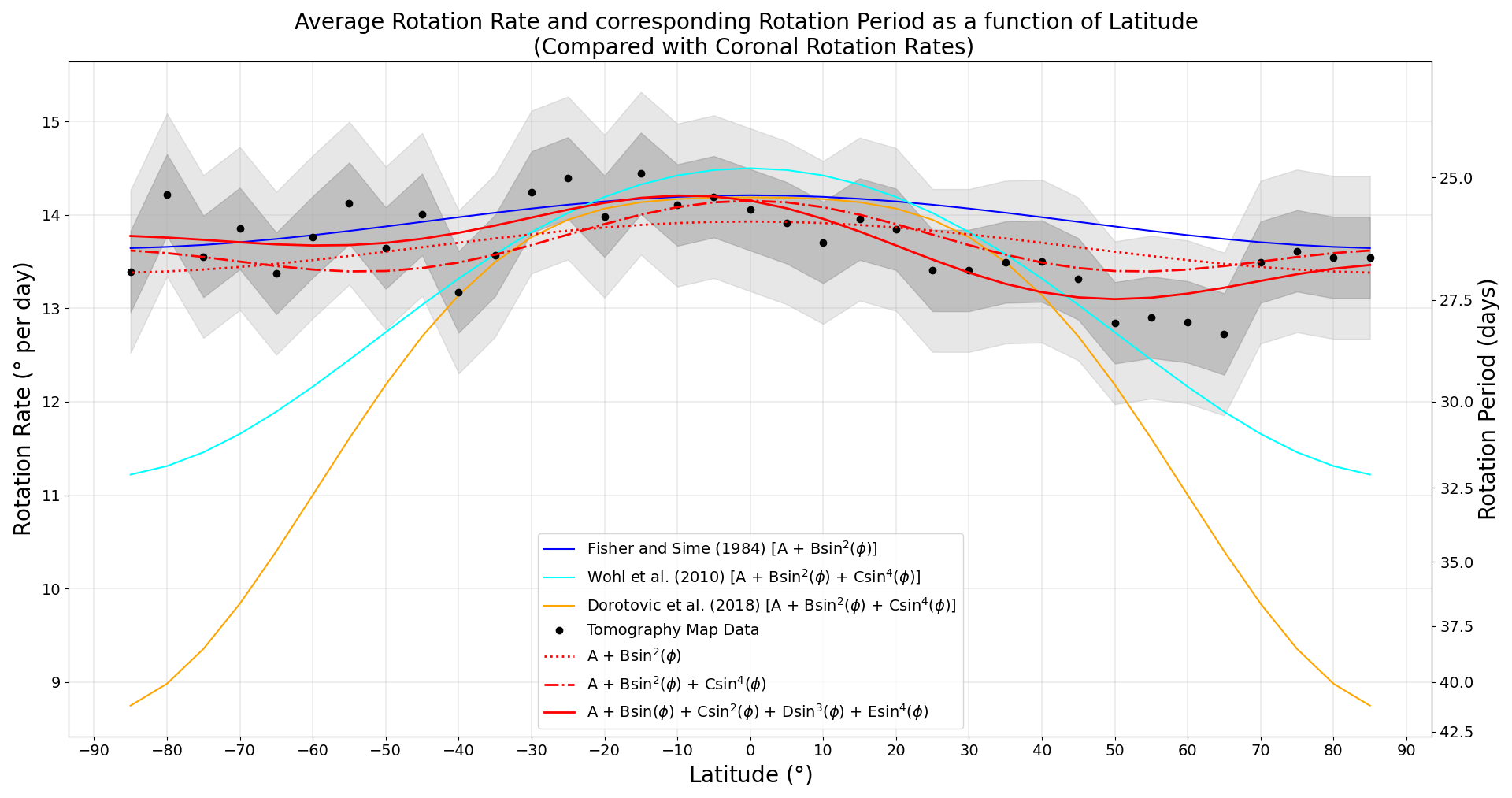}
    \includegraphics[width=0.6\textwidth]{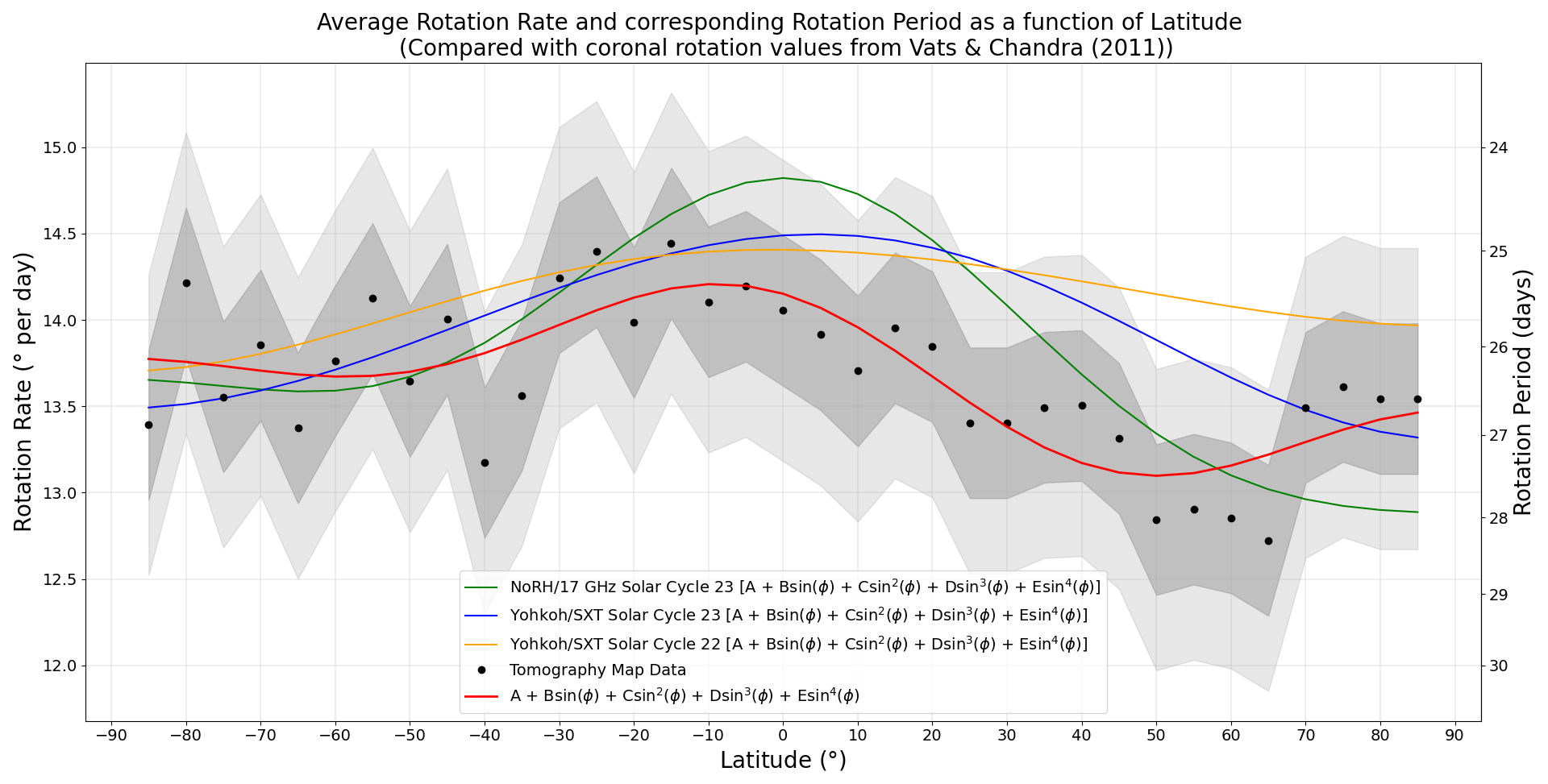}
    \caption{Rotation rates averaged over all time as a function of latitude (black circles), with the standard deviation over time shown by the shaded areas (darker grey and lighter grey for one and two standard deviations, respectively). The dotted red line in the top two panels is a fit to the function of latitude given by Equation \ref{rot1}, and the dashed red line is a fit given by Equation \ref{rot2}. Coloured lines, as indicated in the legend, show photospheric and coronal differential rotation estimates by \citet{1951MNRAS.111..413N}, \citet{1984ApJ...283..373H}, \citet{fisher1984}, \citet{wohl2010}, \citet{2017A&A...606A..72P}, \citet{2018A&C....25..168D} and \citet{2021SoPh..296...25J}. The bottom panel highlights the asymmetry of the rotation by fitting to Equation \ref{mainRot} and comparing the fit with results from \citet{vats2011}.}
    \label{avrot}
\end{figure*}

Figure \ref{avrot} shows rotation rates averaged over all time as a function of latitude, with variances over time shown by the shaded areas. That is, over the whole period of study, the mean and standard deviation of all rotation rates for a given latitude give the mean and standard deviation shown in the plot. These averaged values are fitted to several equations - a fourth-order polynomial function of latitude (Equation \ref{mainRot}) and its simplified, lower-order variants

\begin{equation}
    \label{mainRot}
    \omega = A + B\sin \phi + C\sin^2 \phi + D\sin^3 \phi + E\sin^4 \phi
\end{equation}

Fitting this polynomial to the data gives values for the coefficients as follows: $A = 14.152$, $B=-0.736$, $C=-2.362$, $D=0.584$, and $E=1.839$, which is shown by the solid red lines in Figure \ref{avrot}. The first simplified function used is 

\begin{equation}
\label{rot1}
    \omega = A + B \sin^2\phi, 
\end{equation}

where $\phi$ is the latitude shown as the dotted red line in Figure \ref{avrot}. Fitting our results to this function gives coefficients $A=13.929$ and $B=-0.551$. The second simplified function is 

\begin{equation}
\label{rot2}
    \omega = A + B \sin^2\phi+ C \sin^4\phi,
\end{equation}

as shown with the dashed red line in Figure \ref{avrot}. Fitting our results to this function gives coefficients $A=14.152$, $B=-2.362$, and $C=1.839$. For comparison, several estimates of the photospheric and coronal differential rotation by studies over the past century are also shown \citep{1951MNRAS.111..413N, 1984ApJ...283..373H, fisher1984, wohl2010, 2017A&A...606A..72P, 2018A&C....25..168D, 2021SoPh..296...25J}. This plot emphasises that when rotation rates are averaged over long periods, the resulting means tend to show rigid rotation over latitude. A similar result was shown using tomography by \citet{morgan2011rotation}, and by many other studies based on flux modulation. However, this hides the considerable variance in rotation rates over time (as shown by the shaded areas which show this variance). The corona can, at times, possess considerable latitudinal differential rotation. 

Another significant result from Figure \ref{avrot} is the asymmetry shown between north and south in the bottom panel. The mid-latitudes in the north dip to slow average rotation rates, well below 1\deday\ slower than Carrington, which is not seen in the south. Figure 10a of \citet{morgan2011rotation} shows latitudinal profiles of coronal rotation from the previous solar cycle (1996--2009), and did not find significant asymmetry, although some asymmetry was present when different phases of the cycle were isolated (Figure 10b-d). Several previous studies of coronal rotation have shown a north-south asymmetry \citep{2010A&A...520A..29W,2017A&A...603A.134B,2018AstL...44..727B,2018A&C....25..168D,2021ApJS..252....6H}. \citet{2020MNRAS.499.5442S} found a high asymmetry during the 2011--2014 solar maximum, with less asymmetry in the ascending and descending phases. These results, as do ours, are contrary to \citet{2020A&A...644A..18M}, who suggest that the asymmetry is less pronounced than previously inferred. Our results, which are not based on a long time-series analysis, and instead are based on resolved coronal structure through tomography, show clearly that there is a north-south asymmetry during the recent solar cycle, with a significant slower rotation at northern latitudes of 50\de\ to 65\de.

\begin{figure*}
    \centering
    \includegraphics[width=0.98\textwidth]{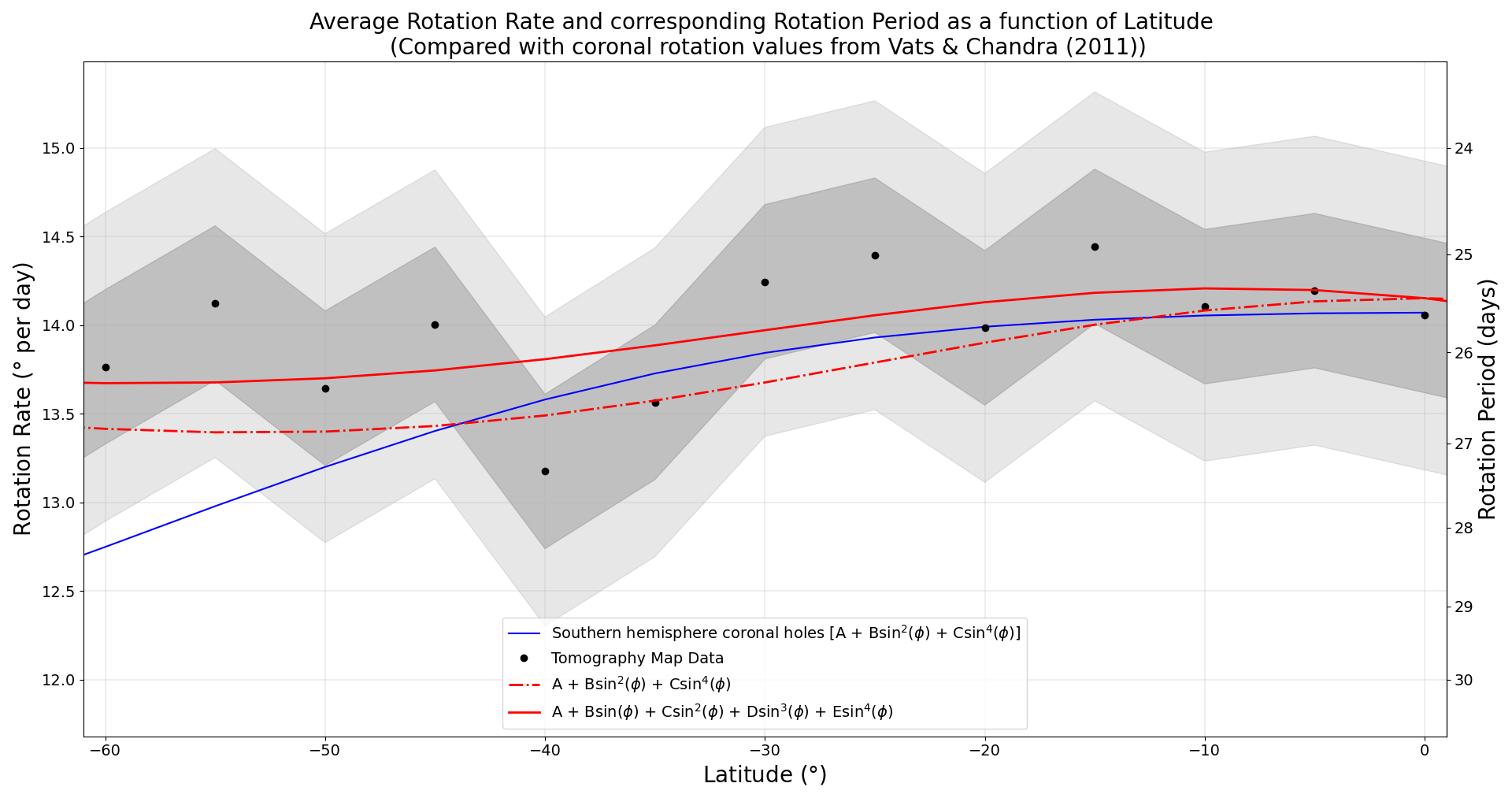} \includegraphics[width=0.98\textwidth]{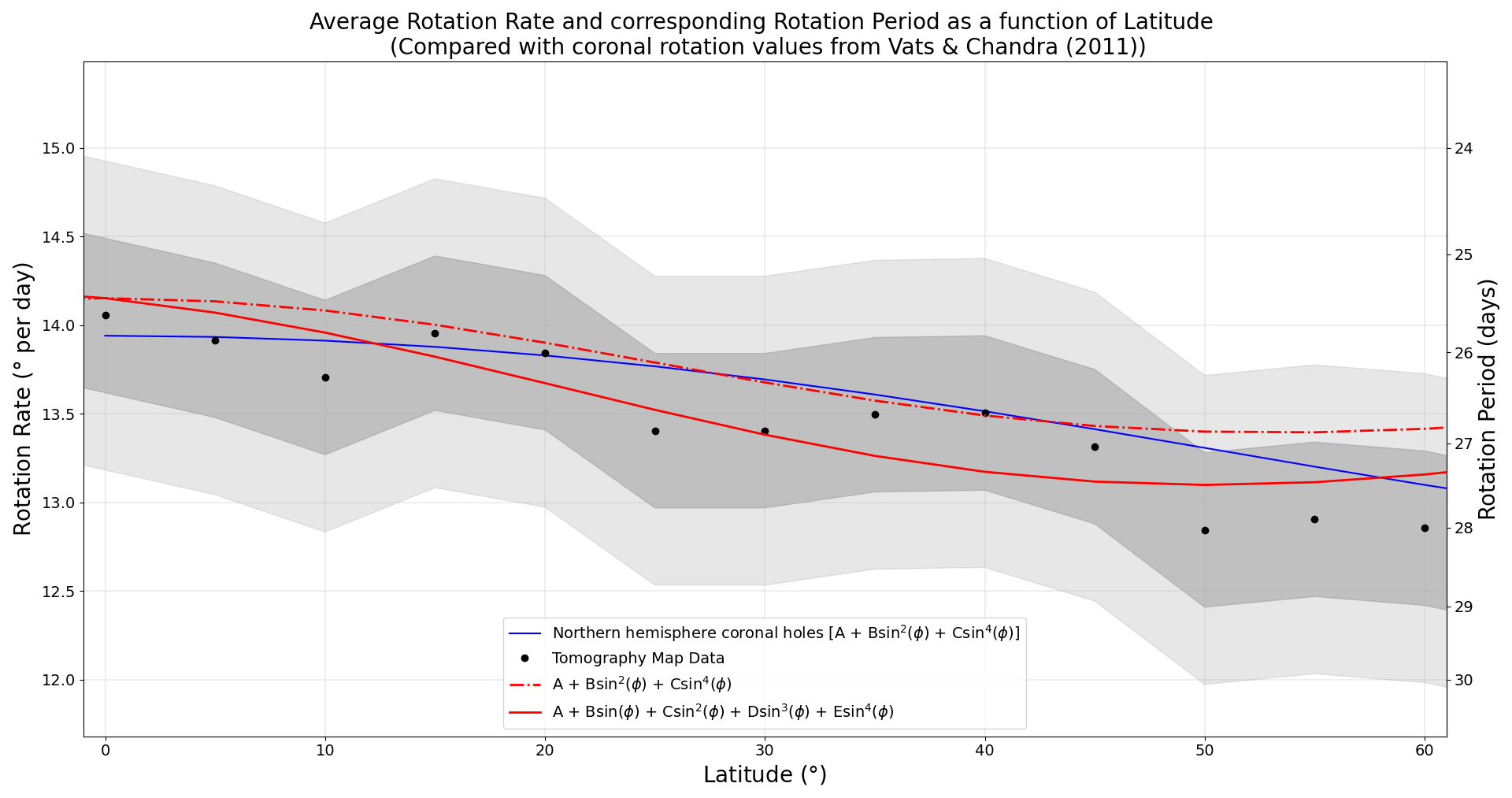}
    \caption{Rotation rates for south (\emph{top}) and north (\emph{bottom}). Our mean rotation rates and variance over time are shown as the data points with shaded areas (darker grey and lighter grey for one and two standard deviations, respectively). The blue coloured lines show estimates of coronal hole rotation from \citet{2017A&A...603A.134B}. The red lines show latitudinal fits to our estimated rates according to Equation \ref{mainRot} (solid red lines) and Equation \ref{rot2} (dashed red lines).}
    \label{fig:bagashvili_comparison}
\end{figure*}

Figure \ref{fig:bagashvili_comparison} shows a detailed comparison between our results and those of \citet{2017A&A...603A.134B}, derived from detailed measurements of coronal holes. The agreement is excellent in the north. In the south, our estimates tend to vary considerably compared to the smooth curve of \citet{2017A&A...603A.134B}, and are significantly higher polewards of $-40$\de. The rotation of coronal streamers must be strongly linked to that of coronal holes, yet the comparison in the south shows significant differences. One major reason may be that high-latitude coronal holes, in the lowest corona, may be measured at extended periods over the solar cycle by \citet{2017A&A...603A.134B}, whereas we can only measure the rotation of high-latitude streamers for a year or two surrounding solar maximum. Another reason may be the highly non-radial structure of the corona, where the measurements of \citet{2017A&A...603A.134B} are made in the lowest corona, and ours are made at extended distances.

\begin{figure}
    \centering
    \includegraphics[width=0.6\textwidth]{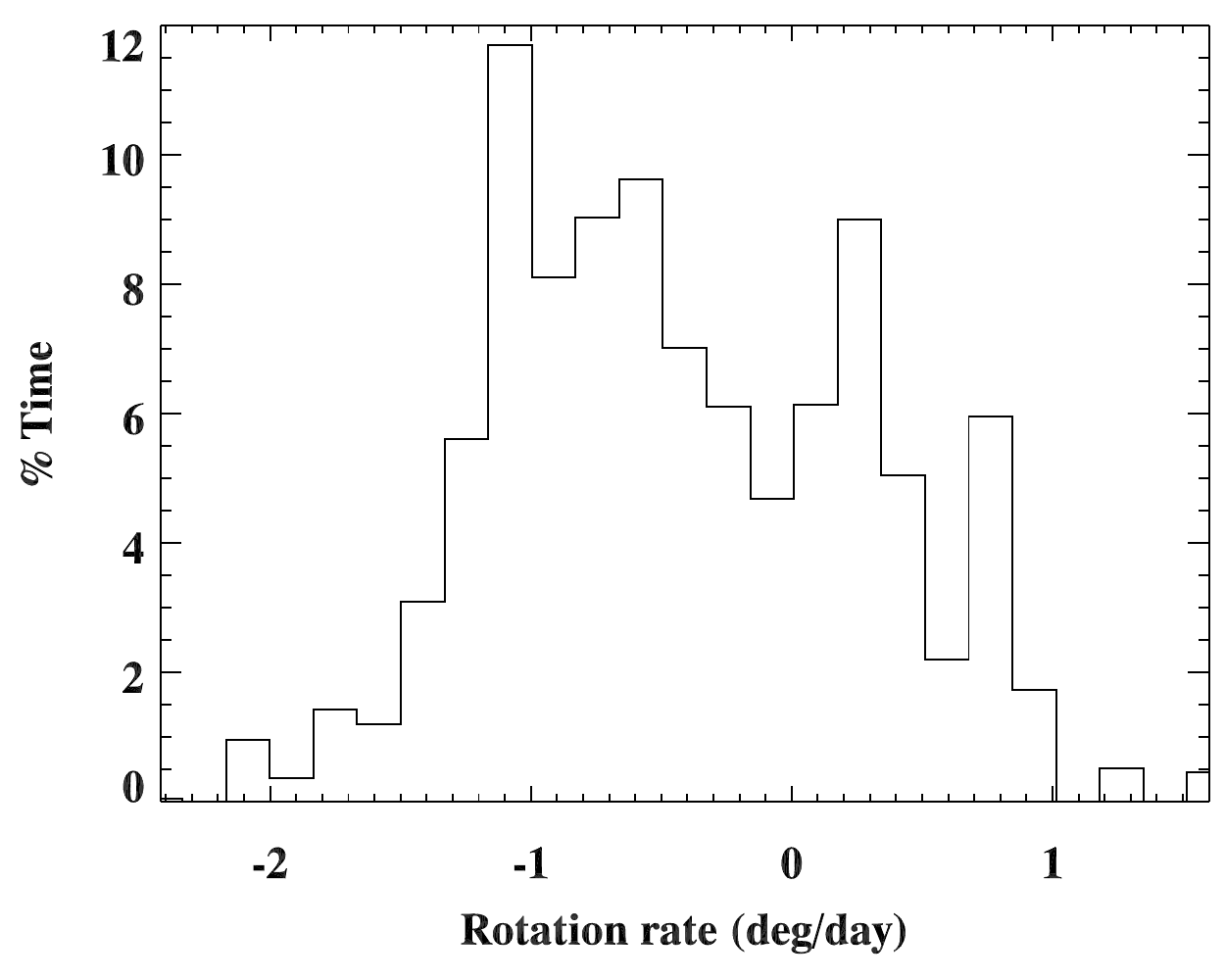}
    \caption{Histogram showing the percentage of time spent at a given rotation rate. The percentage is the fraction of time spanned by streamers measured at that rotation rate relative to the total amount of time spanned by all measured streamers.}
    \label{hist}
\end{figure}

Figure \ref{hist} shows, for all measured streamers, the percentage of time at each rotation rate. The distribution tends towards negative, or slower, rotation rates, with the most probable rate around -1\deday\ relative to Carrington. The bulk of rates are between -1\deday\ and 0.5\deday, although significant time is spent at slower (down to -2.2\deday) and faster (up to 1.6\deday) rates.

%%%%%%%%%%%%%% DISCUSSION %%%%%%%%%%%%%%%%%%%%%%%%%%%%%

\section{Discussion}
\label{discussion}

\subsection{Rotation rates and the latitudes of streamer footpoints}
\label{magmodel}

\begin{figure*}
    \centering
    \includegraphics[width=0.45\textwidth]{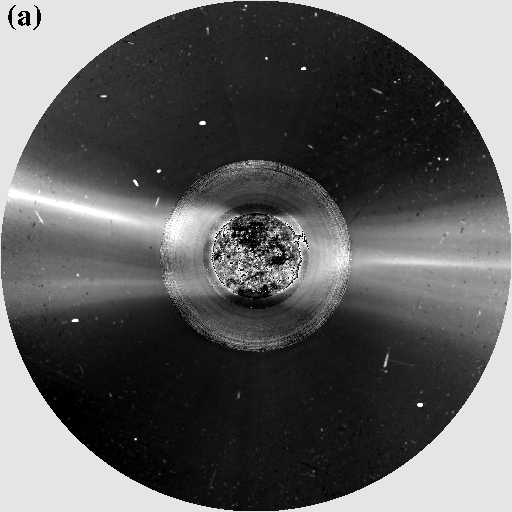}   \includegraphics[width=0.45\textwidth]{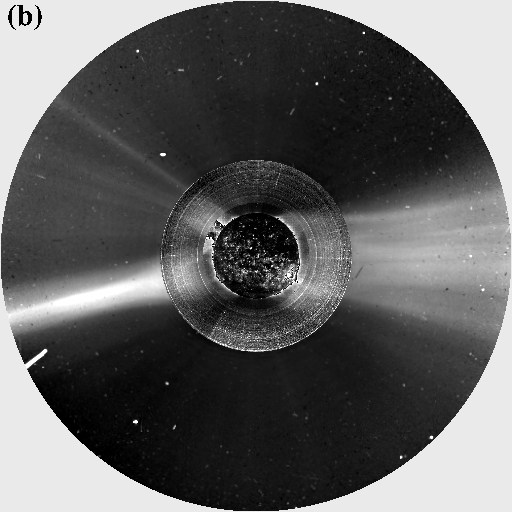}
    \caption{Composite images created using EIT/SOHO 193\AA\ channel data (solar disk and lowest corona), MLSO MK4 coronameter daily average data (middle corona) and LASCO/SOHO C2 data (outer corona) for dates (a) 2009 January 17 and (b) 2010 January 3. Regions above the disk have been processed using a point filter and a normalizing radial graded filter (NRGF, \citet{morgan2005}. }
    \label{composites}
\end{figure*}

A central issue for understanding coronal rotation rates is how the coronal streamers at 4\,\Rs\ are connected to the low solar atmosphere and photosphere. As a simple illustration of this issue, Figure \ref{composites}a shows an image of the 2009 solar minimum corona composed from an EUV observation of the disk and lowest corona by the EUV Imaging Telescope (EIT, \citet{Delaboudiniere1995}) onboard the Solar and Heliospheric Observatory (SOHO, \citet{domingo1995}), a Mauna Loa Solar Observatory (MLSO, \citet{fisher1981}) MK4 coronameter observation of the low corona, and a Large Angle and Spectrometric Coronagraph/SOHO (LASCO/SOHO, \citet{brueckner1995}) C2 observation. If we measure the coronal rotation rate for the most prominent equatorial streamers during this time, the composite image suggests that the footpoints of these streamers encompass a large latitudinal range bridging the equator. The apparent footpoint therefore encompasses a large range of different photospheric rotation rates. For the 2010 ascending phase shown in Figure \ref{composites}b, the prominent streamer between the equator and mid-latitudes in the south-east corona has an apparent narrower footpoint that ranges from south mid-latitudes to just above the equator. Whilst these composite images cannot give us the detailed three-dimensional information needed to link coronal streamers to the low atmosphere, they do serve to illustrate the argument that the connection between streamers in the extended corona and the lower atmosphere is not straightforward, and that we would expect a model of rotation rates based on these connections and photospheric rotation rates to lead to abrupt changes over the solar cycle as the global coronal magnetic field evolves.

\begin{figure*}
    \centering
    \includegraphics[width=0.49\textwidth]{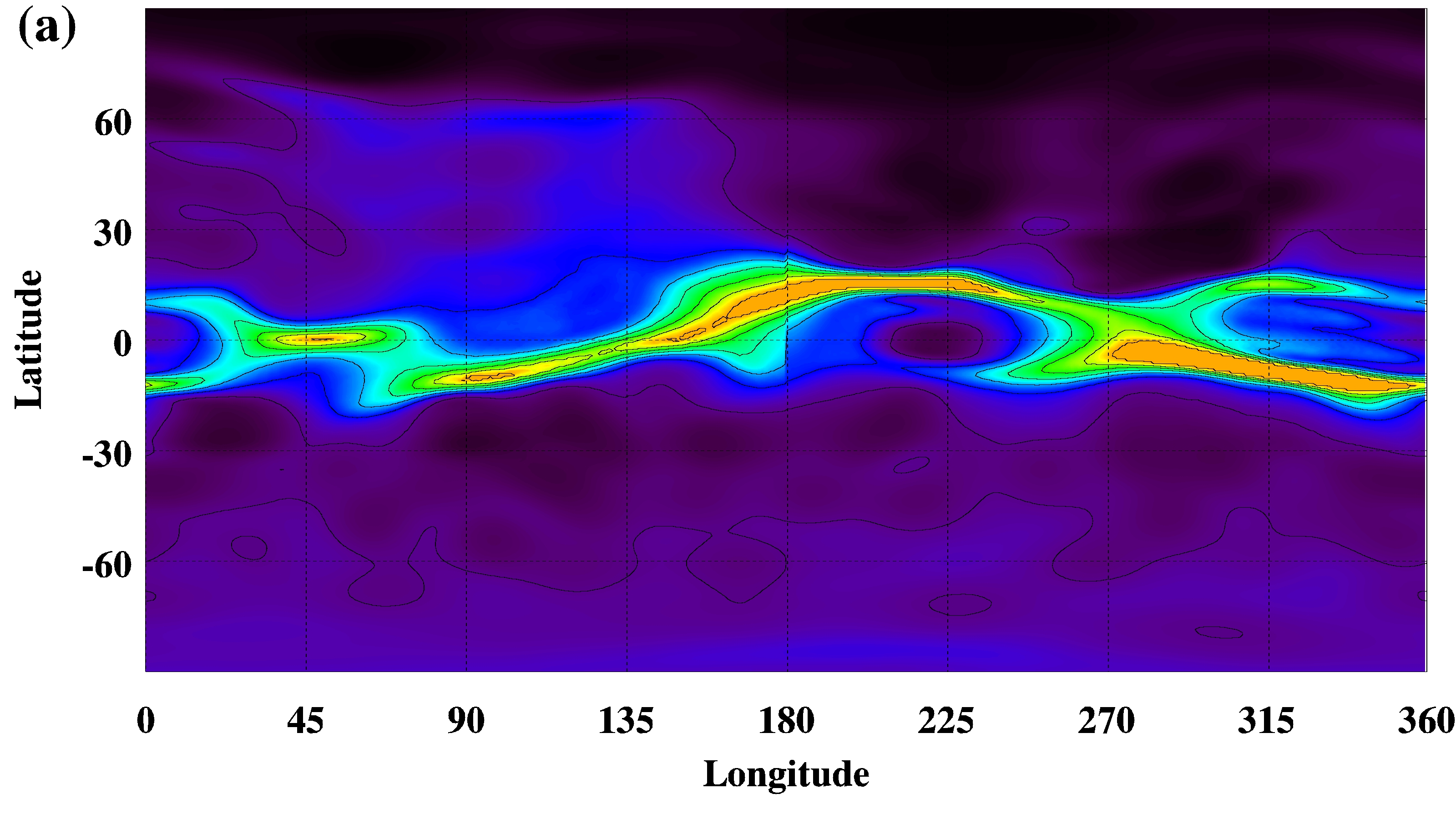}           \includegraphics[width=0.49\textwidth]{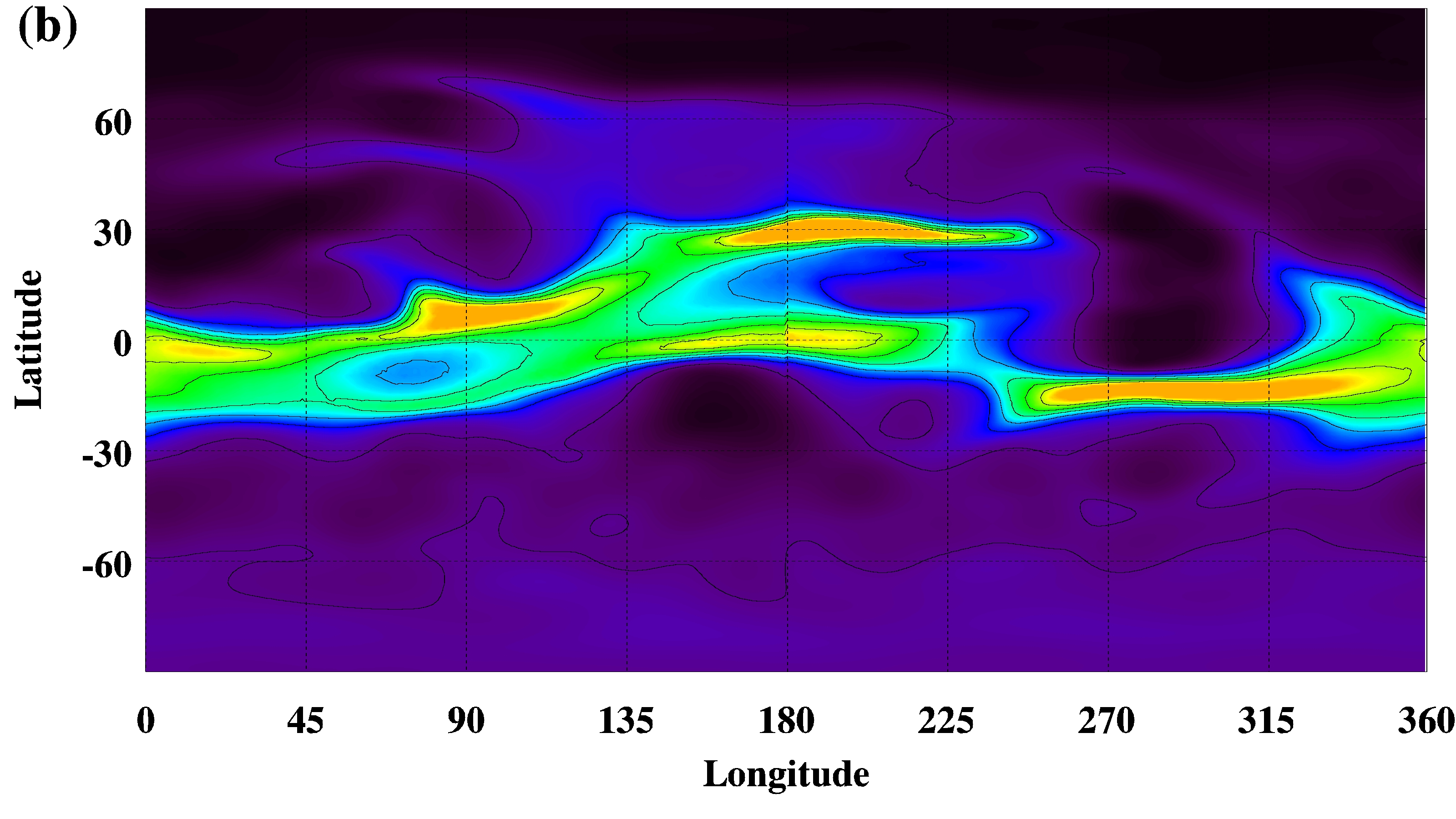}
    \includegraphics[width=0.49\textwidth]{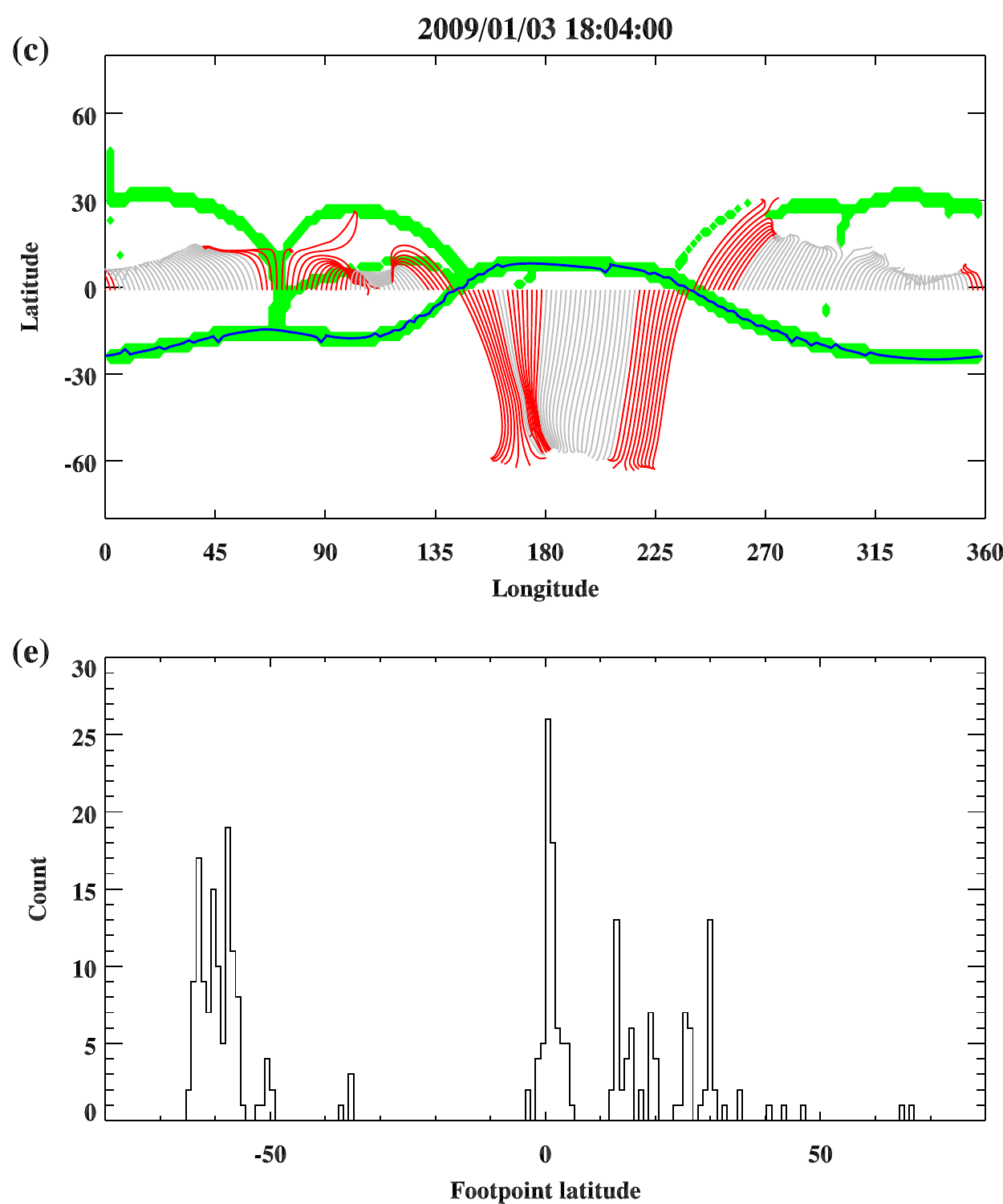}     \includegraphics[width=0.49\textwidth]{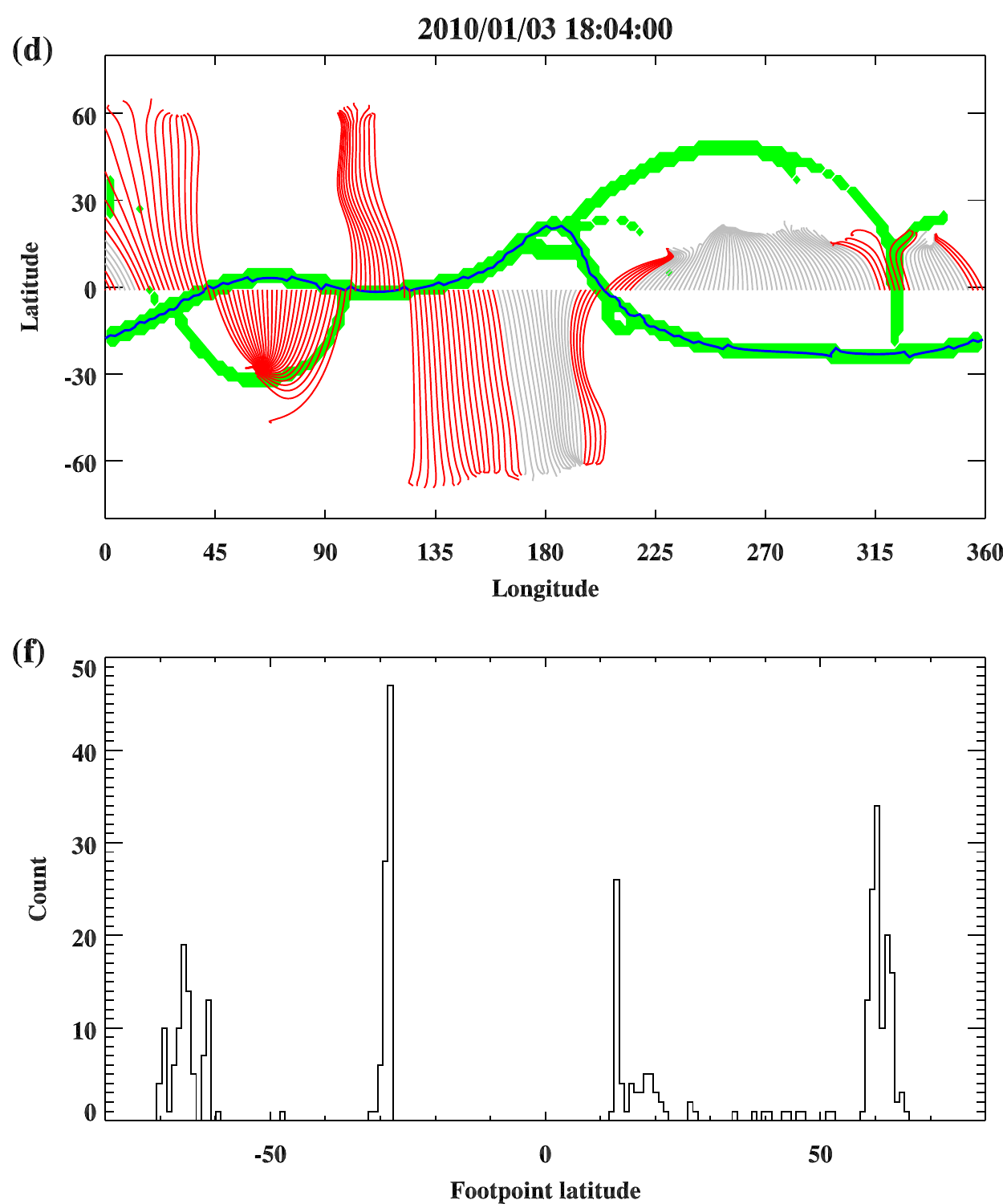}
    \caption{(a) and (d): Tomography density maps for a distance of 4\,\Rs\ for mid-dates 2009 January 3 and 2010 January 3, respectively. (c) and (d): Open field lines from a PFSS model that are near the equator at the source surface distance of 2.5\,\Rs\ are shown as grey or red for dates 2009 January 3 and 2010 January 3, respectively. The green areas shows regions of high convergence where high-density features are expected to reside. The blue line shows the polarity inversion sheet. The red lines are field lines that are close to high-convergence regions at the source surface (that is, both close to the equator at the source surface, and close to a high-density streamer). (e) and (f) For the field lines coloured red in (c) and (d) (i.e. those associated with high-density equatorial streamers), these histograms show the latitudinal distribution of the photospheric footpoints for dates 2009 January 3 and 2010 January 3, respectively.}
    \label{pfss}
\end{figure*}

Figure \ref{pfss} explores photospheric-coronal connections using a Potential Field Source Surface (PFSS) magnetic extrapolation model \citep{newkirk1969, schatten1969}, using the Solarsoft PFSS package based on Helioseismic Magnetic Imager (HMI, \citet{scherrer2012,schou2012})/Solar Dynamics Observatory (SDO) data. The left column of Figure \ref{pfss} shows plots from solar minimum (2009 January) when we measure a positive (a little faster than Carrington) coronal rotation rate near the equator. The right column show plots from a year later (2010 January) when we measure a strong negative rotation rate. The change from positive to negative rate occurs abruptly at the end of 2009, as can be seen in Figure \ref{equatorial} and Figure \ref{rotrat}. Figures \ref{pfss}a and b show the longitude-latitude tomography density maps for 2009 January and 2010 January, respectively. There is no obvious large structural difference between these two periods, only small changes. For example, in 2010 there is a northerly high-density streamer that extends to above 30\de\ north, whereas the 2009 streamer belt is restricted to within $\pm$30\de. Therefore, from inspection of the tomography maps, there does not seem to be an obvious structural change that may be connected to the change in rotation rate.

Figure \ref{pfss} panels c and d show information extracted from the PFSS models for these dates. The green areas are regions of high convergence at the source surface (placed at 2.5\,\Rs). This is a value that quantifies how widely spaced the footpoints of open magnetic field lines are at the photosphere: if neighbouring field lines at the source surface arise from widely separated photospheric regions, the convergence is high. This value, readily derived from PFSS models through field line tracing, was used by \citet{morgan2010structure}, and is equivalent to the squashing factor $Q$ developed in an advanced generalised field mapping method by \citet{titov2007}. High-density coronal streamers (or pseudostreamers) are expected to exist at regions of high convergence, or $Q$. The grey and red lines show field lines that are open at the source surface equator. The red lines are field lines that are close (within 2\de\ in latitude) to a region of high convergence at the source surface (so likely associated with a high-density streamer). We record the location of the photospheric footpoints of these red lines, thus creating a record of the footpoint location of the segments of coronal streamers that are near the equator. The latitudes of these photospheric footpoints are shown in Figure \ref{pfss} panels e and f. Figure \ref{pfss}e, for 2009 January, shows a latitudinal range of approximately -65\de\ to 65\de, with a high count of field lines at the equator, a high count at the extreme south near -60\de, and a more even distribution between the equator and 50\de\ in the north. Figure \ref{pfss}f, for 2010 January, shows a slightly wider distribution between -70\de\ and 65\de. There are two peaks in the distribution near the high-latitude extremities, a peak at -30\de\ and a lesser peak at 15\de. The major difference between the two dates is that 2010 January has no connections to the equator, whereas 2009 January has a strong connection. This gives a possible explanation of the faster rotation in 2009 since the photospheric rotation rate is fastest at the equator. In 2010, the equatorial streamers are linked to higher latitudes where the rotation is slower.

\begin{figure}
    \centering
    \includegraphics[width=0.48\textwidth]{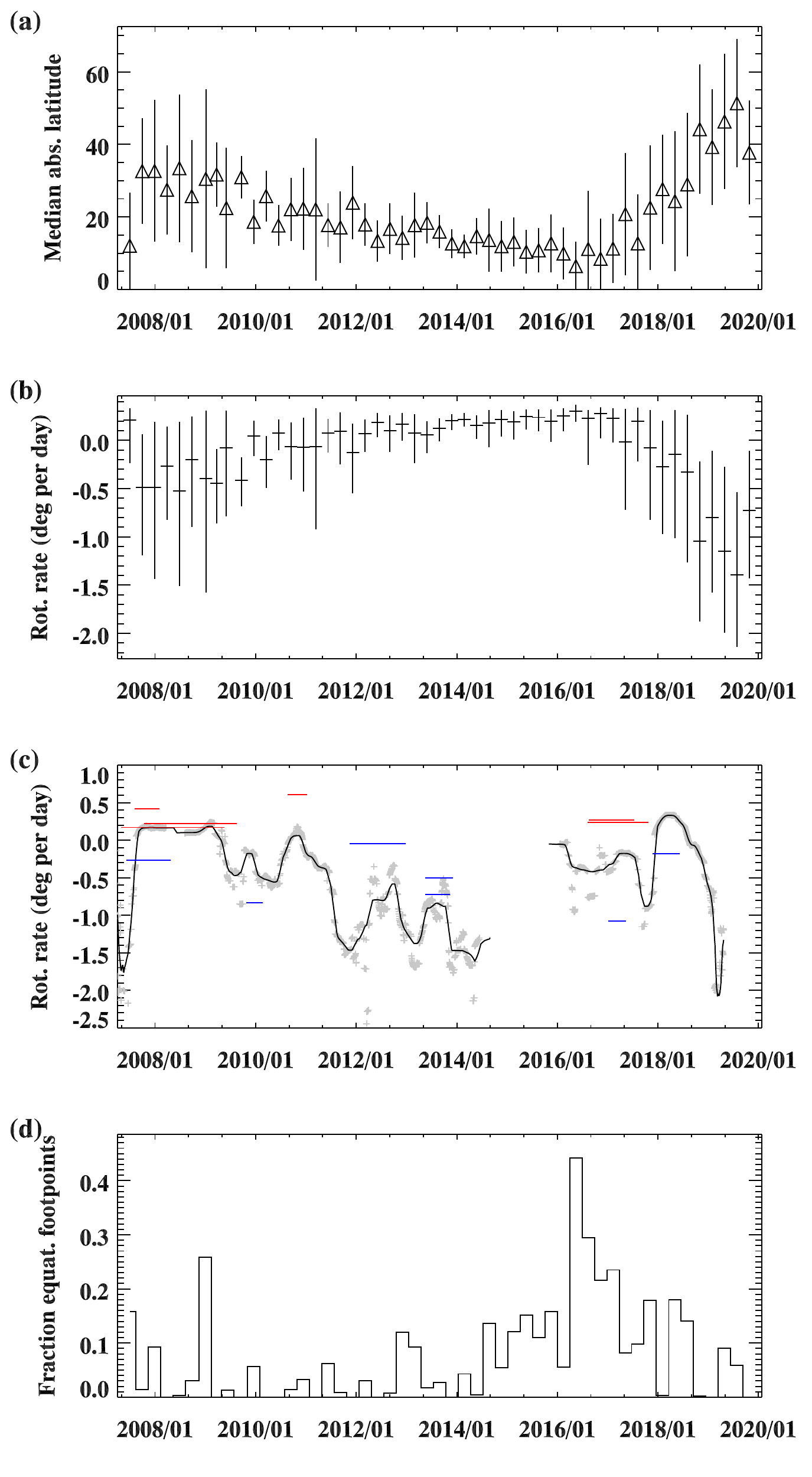} 
    \caption{(a) The median absolute photospheric latitude of field lines associated with equatorial streamers from 2007 to 2020. The error bars show the standard deviation of latitudes from the median for each date. (b) The photospheric rotation rate at the field line footpoints, assuming the sunspot-based rotation rates of \citet{howard1984}. The error bars relate to the variance of latitudes found for each date. The values are relative to the Carrington rate, with zero equal to the Carrington rate. (c) Rotation rates calculated from a correlation procedure (see Appendix \ref{app1}) applied to the equatorial time-longitude tomographical density distribution. The light grey crosses show the unsmoothed rotation rates, and the black line shows the smoothed rates (see Appendix \ref{app1}). The red (blue) horizontal lines show positive (negative) rotation rates (compared to Carrington) gained from the manual measurement of structures in the tomographical maps. (d) The fraction of field line footpoins, relative to the total number of footpoints for that date, that are situated near the equator (within 5\de\ latitude) over time.}
    \label{meanabslat}
\end{figure}

Figure \ref{meanabslat} explores this idea further by applying a similar analysis to the period from mid-2007 to 2020. Every 4 months over this period, we calculate a PFSS model and associated convergence map at the source surface. We extract the field lines that are associated with the equatorial streamers and calculate the median absolute latitude of their photospheric footpoints. These are plotted as a function of time in Figure \ref{meanabslat}a, with the error bars giving the standard deviation of absolute latitudes from the median. The median latitudes are approximately 30\de\ from the equator at solar minimum (2007-2009), dropping to approximately 20\de\ in 2010. There is a gradual decrease to 10\de\ from 2010 to 2016, then a large increase back to 40\de\ between 2016 and 2020. Thus, over the cycle, the segments of coronal streamers near the equator have footpoints spread over a wider latitudinal range, and extending to higher latitudes, at solar minimum. At solar maximum, this range becomes more limited, and moves closer to the equator. 

Based on the photospheric rotation rates of \citet{howard1984}, Figure \ref{meanabslat}b shows the rotation rates at the latitudes of Figure \ref{meanabslat}a. The trend is of course the opposite of the latitudes, with faster (slower) rates at lower (higher) latitudes. These rates are slower than the Carrington rate at solar minimum, increasing to the Carrington rate and slightly above during maximum, then decreasing to very slow rates in the descending phase to the next solar minimum. This is not in agreement with the rotation rates measured by the tomography (see Figure \ref{rotrat}a), where we see small positive rates (relative to Carrington) at solar minimum, and large negative rates after 2010, and a return to positive in 2017. To highlight this disagreement, Figure \ref{meanabslat}c shows equatorial rotation rates gained from a correlation analysis of the tomographical maps (see Appendix \ref{app1} for details of this procedure). Thus the median absolute photospheric latitude of field lines connected to equatorial coronal streamers does not correlate with the equatorial coronal rotation rates - indeed, there is an anti-correlation. Figure \ref{meanabslat}d shows the fraction of equatorial streamer field lines with footpoints within 5\de\ of the equator over the solar cycle. There are prominent peaks of this fraction near 2009 January, and during 2016/2017 that coincide with the most obvious periods of fast coronal rotation. The most obvious long period of slow rotation in 2012 to mid-2014 coincides with a period of generally low fraction of field line equatorial footpoints. This largely qualitative result supports the concept that the rotation rate of coronal streamers are influenced by the latitudinal distribution of their magnetic footpoints near the Sun. If a large fraction of footpoints are near the equator, the streamer is more likely to rotate at the Carrington rate or slightly higher. Note that this does not explain the fast rotation rates of streamers at mid-to-high latitudes. An important point here is that before fast subphotospheric rotations can be used to interpret coronal rotation, we must first understand how and where the corona is connected to the photosphere.

One major uncertainty in this approach is the PFSS model. A comparison of the tomography maps of Figure \ref{pfss} panels a and b with the PFSS convergence values of Figure \ref{pfss} panels c and d show several similarities in the general distribution, but there are considerable disagreements in the actual latitudinal position of structures. This has been shown in previous studies \citep{morgan2010structure, morgan2011longitudinaldrifts}. Despite the usefulness of PFSS to analyse coronal structure, it is a model based on several assumptions (including most importantly the height of the source surface), and this leads to a poorly-quantified uncertainty in the actual position of structures (this is an important consideration given the wide use of PFSS for analysis, solar wind modelling, and interpretation of \emph{in situ} measurements). Another uncertainty in our results is that we have restricted the comparison to equatorial streamers only. At times outside of solar minimum, the streamer sheets meander in latitude, and cross the equator. Our analysis takes only those parts of the streamer sheets that are near the equator. Thus the rotation of an extended streamer sheet may not be dominated by the subset of field lines that are near the equatorial source surface. There are obvious future improvements we can make to this analysis through expanding the scope to streamers at a broader range of latitudes.

\subsection{Implications for solar wind models}
\label{solarwind}
A key parameter for modelling the interplanetary solar wind is the rotation rate of the lower boundary of models. This is a fixed parameter based on the Carrington rotation period (25.38 days sidereal). Our study shows that this parameter should be adjusted according to the estimated rotation rates of streamers over yearly timescales. For a slow wind of 300\kms, the travel time to Earth is approximately 5 days, and a coronal rotation rate difference of -1\deday\ compared to the Carrington rate leads to a systematic error of 5\de\ longitude in the model's output over this travel time period. In the context of short time scales, this is only a minor correction that can be considered negligible in the context of other, larger, uncertainties. However, for analysing solar wind models over the course of longer periods (e.g. a whole rotation or longer), there is obviously a large systematic error that can be corrected given improved rotation rate estimates. 

Therefore, for solar wind modelling of long periods near the solar equator, periods during 2007 to 2009 should remain at the Carrington rotation rate or slightly higher. For periods from 2010 to 2016, a slower rate is recommended, at around -1\deday\ relative to the Carrington rate. From 2017 to 2020, the Carrington rate is appropriate. For future solar wind forecasting, the most recent information on rotation rates available from the tomography maps should be included. This is a service we hope to provide to the community over the coming years, with near-real-time updating of the tomography density maps provided as part of the Space Weather Empirical Ensemble Package (SWEEP) project detailed in the acknowledgements. We also wish to improve the cross-correlation procedure detailed in Appendix \ref{app1}, which provides an automated estimate of the coronal rotation rate.

%%%%%%%%%%%%%% CONCLUSION %%%%%%%%%%%%%%%%%%%%%%%%%%%%%

\section{Conclusions}
Tomographical maps of the coronal density structure give detailed information on the distribution of high-density streamers. In this study, maps from 2007 to 2020 show longitudinal drifts of streamers over time, giving an accurate measurement of coronal rotation rates over a whole solar cycle. The bulk of rotation rates are between -1\deday\ to 0.5\deday\ relative to the Carrington rate, with values measured from -2.2\deday\ to 1.6\deday. Rotation rates can change abruptly at all latitudes, showing that the concept of a rigidly rotating corona (compared to the photosphere) is oversimplistic. The rigid rotation is only found when rotation rates are averaged over long periods, e.g. a whole solar cycle. 

We find a strong north-south asymmetry in rotation rates, with the southern corona rotating more rigidly than the north. Between latitudes of 50\de\ and 65\de\ north we find the consistently slowest rotation rates over the solar cycle. Our results in the north agree well with rotation rates of coronal holes made by \citet{2017A&A...603A.134B}, but disagree at many latitudes in the south. This can be explained by measurements made at different periods during the solar cycle, and with the large difference in height of the different measurements. 

Using the tomography maps and a PFSS magnetic model, we interpret the rotation rates at the equator in terms of the magnetic connection between the coronal streamer belt and the lowest corona. Periods of the cycle with fast equatorial rotation are periods when there are a larger fraction of field lines connected to the equatorial photosphere, and long periods of slow rotation coincide with a period where there is a smaller fraction of equatorial connection. This interpretation explains abrupt changes in the rotation rate in terms of abrupt changes in the coronal structure. For example, as the corona changes from a solar minimum dipole-dominated configuration to a more quadrapolar-dominated configuration in the ascending phase to solar maximum, starting in year 2009, we see an abrupt change from faster to slower rotation at the equator. 

The corona is more likely to be rotating faster than the underlying photosphere at the same latitude. This supports the concept of a subphotospheric influence on large-scale coronal structure, but as highlighted by our analysis of section \ref{magmodel}, the magnetic connection between the corona and the photosphere needs to be better understood. The mechanisms (e.g. interchange reconnection, or more rapid reconfigurations) that allow the corona to rotate faster than the photosphere also need improved understanding, and the use of global simulations such as magnetofrictional models may be useful in this context. Routine measurements of the coronal magnetic field will also be crucial to gain a full understanding.

This study provides improved estimates of the coronal rotation rates that can be adopted by historical solar wind models of the past cycle in order to correct for systematic errors on the order of 5\de\ in longitude, or larger. For equatorial regions, periods during 2007 to 2009 should be slightly faster than the Carrington rotation rate. For periods from 2010 to 2016, a slower rate is recommended, at around -1\deday\ relative to the Carrington rate. From 2017 to 2020, the Carrington rate is appropriate. We plan to provide improved and updated estimates of equatorial rotation rates in the future, as part of the SWEEP project (detailed in the acknowledgements).

\label{conclusions}

%%%%%%%%%%%%%% ACKNOWLEDGEMENTS %%%%%%%%%%%%%%%%%%%%%%%%%%%

\begin{acknowledgements}
We acknowledge STFC grants ST/S000518/1 and ST/V00235X/1, Leverhulme grant RPG-2019-361, and the excellent facilities and support of SuperComputing Wales. STFC project \\ ST/V00235X/1 is the Space Weather Empirical Ensemble Package (SWEEP) project, funded to provide an operational space weather forecasting package for the UK Meterological Office: a collaboration between Aberystwyth University, University of Reading, Durham University, and Northumbria University. The STEREO/SECCHI project is an international consortium of the Naval Research Laboratory (USA), Lockheed Martin Solar and Astrophysics Lab (USA), NASA Goddard Space Flight Center (USA), Rutherford Appleton Laboratory (UK), University of Birmingham (UK), Max-Planck-Institut fu\"r Sonnen-systemforschung (Germany), Centre Spatial de Liege (Belgium), Institut Optique Th\'eorique et Appliq\'uee (France), and Institut d'Astrophysique Spatiale (France). MKIV coronameter data (DOI: 10.5065/D66972C9) courtesy of the Mauna Loa Solar Observatory, operated by the High Altitude Observatory, as part of the National Center for Atmospheric Research (NCAR). NCAR is supported by the National Science Foundation. The SOHO/LASCO data used here are produced by a consortium of the Naval Research Laboratory (USA), Max-Planck-Institut fuer Aeronomie (Germany)), Laboratoire d'Astronomie (France), and the University of Birmingham (UK). SOHO is a project of international cooperation between ESA and NASA. 
\end{acknowledgements}

\appendix
\section{Automated correlation analysis for rotation rates}
\label{app1}

For a given latitude, the tomography data map has dimensions time and Carrington longitude (see, for example, Figure \ref{equatorial} for the equator). The concept for the correlation analysis is to calculate the cross-correlation between a longitudinal slice of the density map at a given time, and a slice at a later time. The maximum peak in the cross-correlation profile gives the longitudinal shift and an estimate of rotation rate. This can be applied to multiple time steps throughout the map to give a time series of estimated rotation rates. In practice, noise and errors in the tomography map, plus actual structural changes in the corona, lead to high variations and incoherence in the resulting rotation rate time series. To gain meaningful results the following steps are made:
\begin{itemize}
    \item Prior to cross-correlation, the time-longitude map is smoothed along the time direction over 11 time bins, or approximately 16 days.
    \item The cross-correlation is made for 5 selected time intervals: 30, 55, 82, 108, and 135 days. Note that results for the final 135 days of the time series are discarded, corresponding to the longest time interval for cross-correlation. 
    \item For each of the 5 time intervals, the peak cross-correlation is recorded at each time step, and the longitudinal lag of the peak is converted into a rotation rate based on the longitudinal lag and time interval. 
    \item For each of the 5 time intervals, the time series of lags is smoothed with a median sliding window of width 125 days.
    \item The final rotation rate value is calculated as an average of the 5 values given by the 5 time intervals. 
    \item For further analysis, a sliding window smoothing is applied to the rotation rate time series. For the example shown in Figure \ref{meanabslat}c, we median-smooth initially with a sliding window width of 118 days, then average-smooth over a width of 44 days. The choice of these widths provides a reasonably coherent and smooth time series of rotation rates.
\end{itemize}

\begin{figure}[h!]
    \centering
    \includegraphics[width=0.55\textwidth]{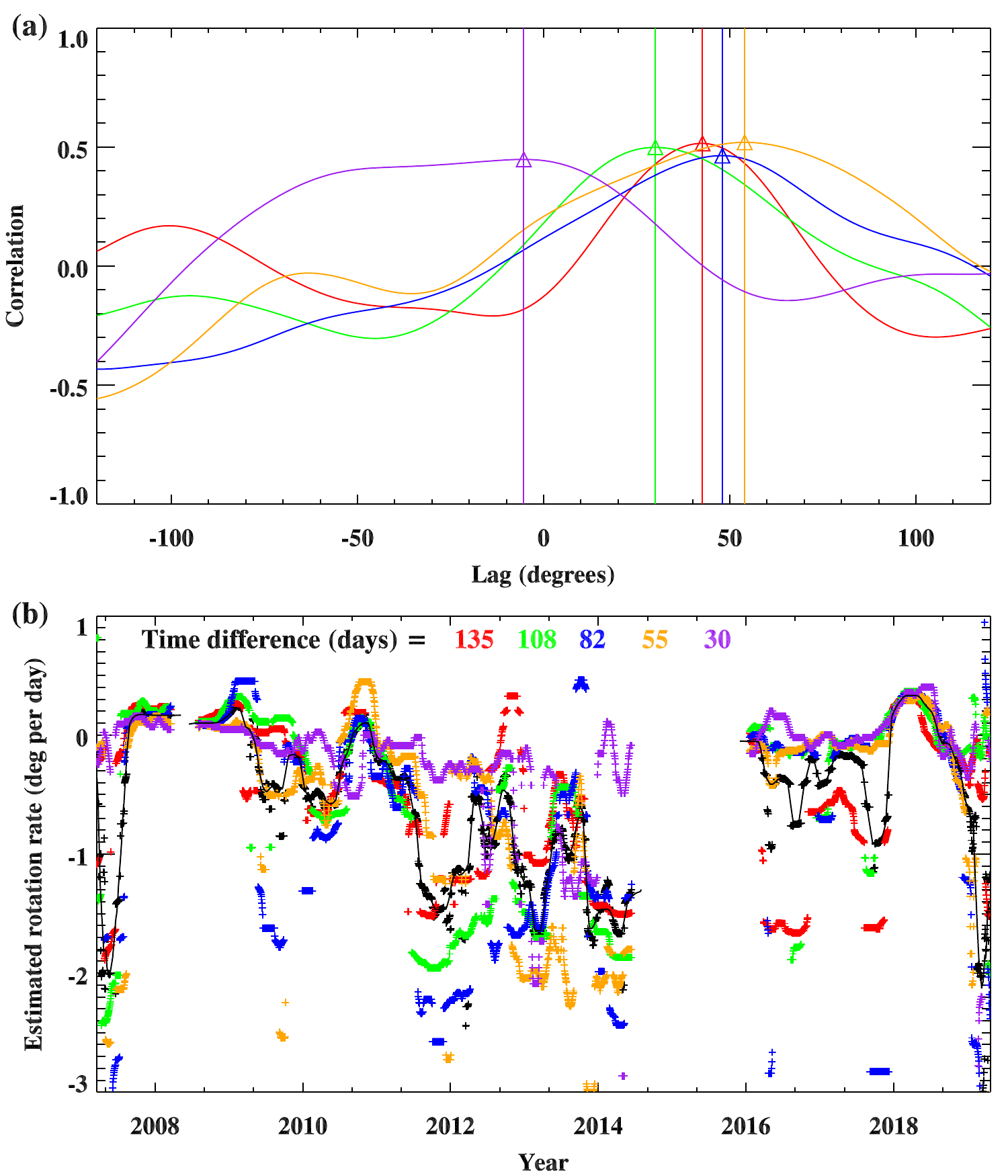} 
    \caption{(a) Example of cross-correlation curves between the equatorial densities for base date 2009 April 7. The different colours show the cross-correlation with densities measured at subsequent dates 30 (purple), 55 (orange), 82 (blue), 108 (green), and 135 (red) days after this date, as shown in the legend to part (b). The point of maximum cross-correlation is indicated by the triangles, with the vertical lines showing the corresponding longitudinal lag in degrees. (b) The estimated rotation rates, in degrees per day relative to the Carrington rate (Carrington rate is zero), as a function of time as calculated for each choice of time interval. The black points and line show the mean over all 5 time intervals.}
    \label{appfig}
\end{figure}

Figure \ref{appfig}a shows an example set of cross correlation curves, as a function of longitudinal lag, for date 2009 April 7. This date corresponds to the base date, with cross-correlations calculated between the base date and the 5 subsequent time intervals. This example shows the difficulty of automatically estimating rotation rates. The peaks are broad, which reflects high uncertainty in choosing the lag at the point of maximum cross-correlation. Often two or more peaks are seen, which shows that different density features may be moving at different rates, or that there are changes in density not associated with rotation. For this example, the 30-day interval is at a negative lag, whilst the other 4 intervals are tightly grouped at around 50\de\ lag. For the optimal detection of rotation, the lags should all be the same sign, and should also increase in magnitude linearly with increasing time interval - this is clearly not the case for this example, and is a further reflection of the difficulties involved in this kind of automated analysis.

Figure \ref{appfig}b shows the estimated rotation rates as a function of time for each choice of time interval. There are times when the 5 time intervals agree well (for example, years 2008-2009), and this is probably due to the low activity of the sun during solar minimum. At this time, changes in the configuration of the coronal density structure are slow, and the method can more reliably detect the dominant rotation rates. At other times, values can vary widely, particularly during solar maximum in years 2012 to 2014. Despite this variation, all five time intervals show a reduced rotation rate. The 30-day interval shows only small deviations from zero throughout the whole period.

%%%%%%%%%%%%%% BIBLIOGRAPHY %%%%%%%%%%%%%%%%%%%%%%%%%%%
\bibliography{biblio.bib}

\end{document}